\RequirePackage{fix-cm}
\documentclass{svjour3}                     
\smartqed  
\usepackage{graphicx}
\usepackage{subfig}             
 \usepackage{threeparttable}
\usepackage{mathabx}
\usepackage{oldgerm}
\begin{document}

\title{Coupled Orbit-Attitude Dynamics of High Area-to-Mass Ratio (HAMR) Objects: Influence of Solar Radiation Pressure, Earth's Shadow and the Visibility in Light Curves
}


\author{Carolin Fr\"uh          \and
        Thomas M. Kelecy        \and Moriba K. Jah}


\institute{C. Fr\"uh  \at
              NRC Postdoc, Air Force Research Laboratory, Space Vehicles Directorate, Research Assistant Professor, University of New exico, Albuquerque, USA \\
              Tel.: +1-505-277-2761 \\
              Fax: +1-505-277-1571\\
              \email{carolin.frueh@gmail.com}           
           \and
           Thomas M. Kelecy \at
              The Boeing Company, 5555 Tech Center Dr., Ste 400, Colorado Springs, CO 80919
	\and 
	Moriba K. Jah \at Air Force Research Laboratory, Space Vehicles Directorate, NM, 87117, USA
}

\date{Received: 8 January 2013 / Accepted: 20 August 2013, Preprint. Final version published in Celest Mech Dyn Astr (2013) 117:385–404, DOI 10.1007/s10569-013-9516-5}

\maketitle

\begin{abstract}
The orbital and attitude dynamics of uncontrolled Earth orbiting objects are perturbed by a variety of sources. In research, emphasis has been put on operational space vehicles. Operational satellites typically have a relatively compact shape, and hence, a low area-to-mass ratio (AMR), and are in most cases actively or passively attitude stabilized. This enables one to treat the orbit and attitude propagation as decoupled problems, and in many cases the attitude dynamics can be neglected completely. The situation is different for space debris objects, which are in an uncontrolled attitude state. Furthermore, the assumption that a steady-state attitude motion can be averaged over data reduction intervals may no longer be valid. Additionally, a subset of the debris objects have significantly high area-to-mass ratio (HAMR) values, resulting in highly perturbed orbits, e.g. by solar radiation pressure, even if a stable AMR value is assumed. Note, this assumption implies a steady-state attitude such that the average cross-sectional area exposed to the sun is close to constant. Time-varying solar radiation pressure accelerations due to attitude variations will result in un-modeled errors in the state propagation. This work investigates the evolution of the coupled attitude and orbit motion of HAMR objects. Standardized pieces of multilayer insulation (MLI) are simulated in a near geosynchronous orbits. It is assumed that the objects are rigid bodies and are in uncontrolled attitude states. The integrated effects of the Earth gravitational field and solar radiation pressure on the attitude motion are investigated. The light curves that represent the observed brightness variations over time in a specific viewing direction are extracted. A sensor model is utilized to generate light curves with visibility constraints and magnitude uncertainties as observed by a standard ground based telescope. The photometric models will be needed when combining photometric and astrometric observations for estimation of orbit and attitude dynamics of non-resolved space objects.
\end{abstract}

\section{Introduction}
\label{intro}
In 2004  a new population of space debris objects in GEO-like orbits was detected by T. Schildknecht \cite{Schildknecht03,Schildknecht04a}.The observations of these objects could only be fitted in an orbit determination, when estimating a very high area-to-mass ratio significantly larger than $1\,m^2/kg$. The orbits of those high-area-to mass (HAMR) objects are highly perturbed, in contrary to low area-to-mass ratio objects, especially by non-conservative forces. In case of objects in near-geostationary orbits, non-conservative perturbations are dominated by the effects of solar radiation pressure. The orbital evolution of those objects differs significantly from those of other uncontrolled objects with low area-to-mass ratio (AMR). Simulations of objects with high but constant AMR in GEO-like orbits revealed a dominant period of one nodal year in the inclination. The amplitude of the inclination variations is proportional to the magnitude of the AMR value. Those orbits also show periodic significant changes in the eccentricity \cite{Liou2005,anselmo,valk1,valk2}. \\
\\
The long term evolution of the orbits, which were determined with the real observations and for which high AMR values were estimated, are within the uncertainties and the simplifications implied in the propagation models in agreement to the theoretical predictions, but also indicated that the AMR value might not to be stable over time \cite{ichhamr}. The comparison of estimated AMR values and measured spectra of HAMR objects with the AMR and spectra measured in labs of space materials, used in satellites, suggest that at least some of the HAMR objects could be pieces of multi layer insulation (MLI), peeled off aging satellites \cite{thomasIAC08,thomas_spec_IAC10,cowardinamos09,Joergensen_spec_AMOS03,Abercomby_AMOS09}.  \\ 
\\
In contrast to objects with low AMR values, a change in the AMR value has significant influence on the orbital evolution. Therefore, the orbital motion cannot be fully decoupled from the attitude propagation. Simple approximations, such as assuming a stable attitude motion around one axis  or that the attitude motion is averaged out over longer time periods, may not be adequate for accurate predictions necessary for re-acquisition of observed objects. \\
\\
In the work presented here, the coupled orbit and attitude propagation of HAMR objects in near geosynchronous orbits are investigated. Different sheet like HAMR objects are simulated and the influence of the gravitational field, direct solar radiation pressure and Earth's shadow. Orbit and attitude motion is studied. The material properties of MLI are simulated, and the objects are assumed to be rigid bodies. Simulated light curves, derived from typical bidirectional reflectance distribution functions (BRDF) that are combined with the dynamic attitude time history, are generated to illustrate how the dynamics are manifested in those measurements. A sensor model is applied to simulate the measurement taken by a ground based optical sensor. All results of the sheet like HAMR objects with a fully coupled attitude orbit propagation, are compared to canon ball models with the same averaged reflection properties and the same effective area-to-mass ratio. \\
\\
The method is very accurate and allows to understand the dynamic evolution of the objects, but also computationally demanding. Once this understanding is provided, several simplification models can be employed, which save computational time and allow for longer propagation times. A common simplification is the canon-ball model, which assumes a constant area-to-mass ratio, but as the work presented here shows, this is not a adequate assumption, neither for nearly stable nor highly spinning objects, because the naturally induced spin rates are not uniform. However, this assumption can be useful in the detection and interpretation of general trends of whole object populations, when propagation intervals of several years are assumed. A better approximate solution for shorter time intervals may be obtained using, e.g. a semi-coupled integration approach \cite{ich_semi}.
\section{Simulation of the Objects}
\label{obi}
\subsection{Background}
In common satellites, MLI is used to thermally insulate the satellite bus. Up to 20 and more MLI layers are used. The two outer layers have a thickness of 0.5 to 1.0$ mm$, whereas the inner layers a thickness of 0.25$mm$. Those materials have densities of 1.4 to 2.2$g/cm^2$, which translates to mean area-to-mass ratios of around 6$m^2/kg$ for the outer MLI layers and around 110$m^2/kg$ for the inner layers , respectively. Typical materials include Kaptons$^{\textregistered}$, Mylar$^{\textregistered}$, Teflon$^{\textregistered}$ FEP, PET and Tedlar$^{\textregistered}$ PVF \cite{sven_mli,redbook}. The MLI layers peal off the satellite over time, equivalent solar hours are the dominant factor for material deterioration. Returned samples of Aluminum-FEP of the Hubble space telescope showed cracks in regions of constrained loading. Those samples also showed that the layers have the tendency to curl up during pealing off \cite{dever1998mechanical}. The fragmentation of MLI has been investigated in various fragmentation tests \cite{liou2009}. The size of the single pieces produced in such a fragmentation range from  many pieces with centimeter size up to larger pieces having an edge length of up to half a meter. The materials are normally coated, most of the times with aluminum, either on both or one side, leading to albedo values of 0.95, but investigations also showed that the reduction in reflectivity because of proton and electron fluxes can be as high as 0.2\cite{sven_mli}. \\

\subsection{Object Shapes}
Two different object shapes have been simulated. Flat rigid sheets and spherically shaped canon balls.The flat rigid sheets serve as models for our actual HAMR objects, whereas the canon ball object serve as reference objects. \\
\begin{figure}[ht!]
  \centering
\includegraphics[ width=0.55\textwidth]{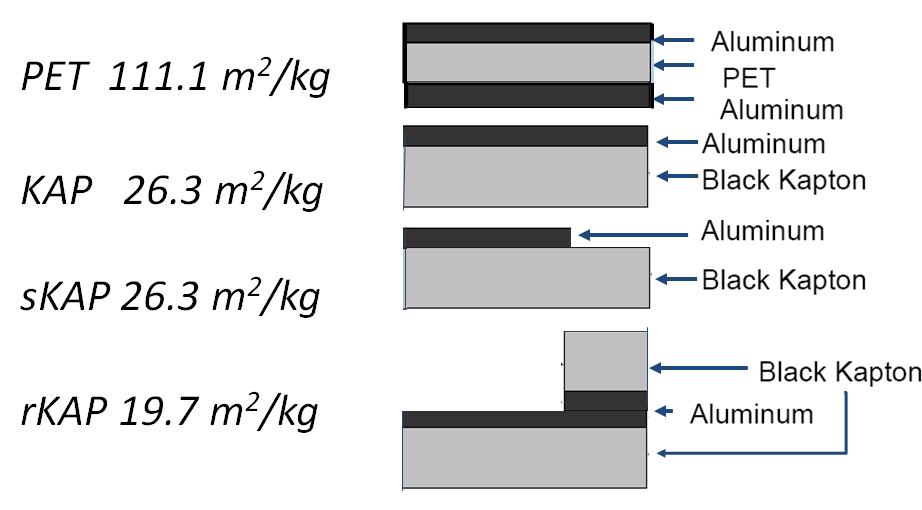}
  \caption{\bf Different flat plate objects and AMR values (scematic picture). All objects are simulated as flat rigid plates.}
  \label{simi}
\end{figure}
\\
The sheets are modeled to be flat, rigid and of an area of one square meter. Four different objects are simulated, consisting of the two materials, double coated PET and single coated Kapton, with the reflection values of  $C_{s,d,a}=0.60;0.26; 0.14$ for the coated side and of $C_{s,d,a}= 0.00;0.10;0.90$ for the uncoated side. Details can be found in the supplement materials. The objects are illustrated in Fig.\ref{simi}. The first object type, called \textit{PET} is a single sheet of PET material, coated on both sides. The second object is a single coated Kapton sheet, labeled \textit{KAP}. The third object is a single coated Kapton sheet, where the coating of the coated side is not complete any more but eroded in the harsh space environment over time. This means that one third of the object surface has the same reflection properties than the back side, and two thirds still the coated reflection properties. Coating can erode either due to long-time sun exposure, or by the acrylic transfer film, which holds together different layers of MLI in a satellite \cite{dever1998mechanical}. The adhesive material is placed selectively, and alternated by stitching. The material is most likely to break under stress at the stitched places. Hence it is assumed that coating came off on one side, covering about one third of one side of the surface, opposite to the stitched side where the material first shed, is a reasonable assumption. This object is labeled \textit{sKAP}, see Fig.\ref{simi} for illustration. The forth object type is a single coated Kapton sheet whose surface is one third larger than the previous objects. This exact part has curled up and folded back on the front side. This has two effects: for one the AMR value of the object has increased and the center of mass is not identical with the center of pressure any more. Furthermore it has the same reflection properties as the object \textit{sKAP}. It is labeled \textit{rKAP}. The four flat sheet object types are illustrated in Fig.\ref{simi}, as well as the AMR values are given. All objects, also \textit{rKAP}, are simulated as flat plates, because the lateral extension of the object (one one meter) is much larger than the thickness of the object even when double layered (1 mm). In the course of the paper, those objects are simulated in a full six degree of freedom propagation. As a comparison  simulations are performed in which the attitude state is enforced to be fixed to the initial values. In the latter case the objects are labeled with an additional term $stable$ after the object name.\\
\\
The canon ball objects serve as reference objects, they do not represent realistic objects but serve as a reference in order to represent the standard model, which is generally used for all object shapes in the three dimensional orbit propagation of filter parameter estimation, neglecting more complex shapes and eventual attitude motion. Four different kinds of canon ball reference objects have been simulated, in accordance with the four different sheet object types. The first reference object is an object with the same AMR value and same same reflection properties as the PET object, it is labeled \textit{ball-PET}. This corresponds to the physical equivalent canon-ball shaped object. However, the initial angle that the plate towards the sun has, is about 77.8 degrees, which leads to  an effective AMR of 25.40 $m^2/kg$ only. This would also be the value when estimating the AMR from a short observation arc, covering few minutes at the beginning of the propagation interval, under the assumption of a canon-ball model in the filter estimation process, it is labeled \textit{ball-PETeff}. The third and the forth object are labeled \textit{ball-sKAP} and \textit{ball-rKAP}, which have the same AMR values and the averaged reflection properties as corresponding flat plate objects, respectively. A list of the objects can be found in the supplement materials.

\section{Orbit and Attitude Propagation}
The geocentric equations of motion for an object are described as follows:
\begin{eqnarray}
\ddot{\vec x}=-GM_{\Earth} \nabla V(\vec x)-G\sum_{k=1,2}M_k\left[\frac{\vec x-\vec x_k}{|\vec x-\vec x_k|^3}+\frac{\vec x_k}{x_k^3}\right]+\sum_{l} \vec a_l
\label{11}
\end{eqnarray}
where $\vec x$ is the geocentric position of the object, $G$ the gravitational constant, $M_{\Earth}$ the Earth mass and $V(\vec x)$ the Earth gravitational potential. For its representation the formulation of Pines \cite{pines} was chosen, and transformed in the Earth centered space fixed coordinate system. The third body gravitational perturbations of the Sun and Moon (k=1,2)  with the states $\vec x_k$ have also been taken into account. Finally, $\sum \vec a$ is the sum over all non-gravitational accelerations acting on the satellite.  These latter perturbations can include accelerations due to direct solar radiation pressure (SRP) and, hence, the attitude dependence finds its way into the equations of motion.  If attitude dynamics are modeled, these must be included in the integration of the equations of motion.

\subsection{Spherical Body}
For a spherical body with uniform reflection properties attitude motion does not have any effect on the orbital evolution. For the canon ball object, a sphere with radius $r$, the direct radiation pressure acts as a perturbing force on the orbit, and takes the following form:
\begin{eqnarray}
\vec F_{\mathrm{rad}}=-\frac{\mathrm{4\pi r^2}}{m}\frac{E}{c}
\frac{A_{\Earth}^2}{|\vec x-\vec x_{\Sun}|^2}\cdot \big (\frac{1}{4}+\frac{1}{9}C_{\mathrm{d}})\hat{\vec S}
\label{rad}
\end{eqnarray}
$m$ is the total mass of the object, $E$ is the solar flux, $A_{\Earth}$ the astronomical unit, $\vec x_{\Sun}$ the geocentric position of the sun, $c$ velocity of light, $\vec S$ the direction of the radiation source, $\vec x$ is the position vector of the object, and $C_d$ is the diffuse reflection coefficient.

\subsection{Non-spherical Body Consisting of Flat Surfaces}
For objects, which are not uniform and spherical, attitude motion is relevant. The dynamic equations of motion a rigid body can be calculated using Euler's equations \cite{Wertz}, where the body is approximated as sum over the $n$ facet- and volume elements approximating the body's shape and mass, respectively:
\begin{eqnarray}
&\frac{d}{dt} (I\vec \omega)=\sum_{l=1}^m \vec \tau_{l} - \vec \omega\times (I\vec \omega) \\\nonumber
&\mbox{ where:}\hspace{0.5cm}I_{\alpha\beta}=\sum_{k=1}^n m_{k}(r_k^2\delta_{\alpha\beta}-z_{k,\alpha}z_{k,\beta}), \hspace{0.1cm} r_k=\sqrt{\sum_{\alpha,\beta,\gamma}z^2_k}
\label{1}
\end{eqnarray}
where $\vec \omega (t)$  is the angular velocity body rate. $I_{\alpha\beta}$ is the tensors of the net moments of inertia of each volume element, with mass $m_k$ located at position $z_{\alpha\beta\gamma}$, $\delta$ is the kronecker delta. The moments of inertia are constant over time, under the assumption of a rigid body. $\sum_{l} \vec \tau_l$ represents the sum of the disturbance torques. In the case of space debris objects no control torques are present. The vectors are represented and time-derivatives taken in a body-fixed reference frame. The kinematic equations can be represented in terms of attitude as quaternions \cite{Wertz}:
\begin{eqnarray}
\frac{d \vec q}{dt}=0.5\cdot\Omega\cdot\vec q 
\hspace{1cm}\mbox{ with:}\hspace{0.5cm}
\Omega= \left [ \begin{array}{cccc}
0 & \omega_3 & -\omega_2  & \omega_1\\
-\omega_3 & 0 & \omega_1 & \omega_2\\
\omega_2 & -\omega_1 & 0 & \omega_3\\
-\omega_1 & -\omega_2 &-\omega_3 & 0\end{array} \right ]
\end{eqnarray}
The gravitational torque is approximated assuming that the distance between the geocenter and the geometric center of the object $\vec x$ is much larger than the extension of the object itself for each volume element of the object. It can be expressed as the following with the moments of inertia tensor defined in Eq.\,\ref{1}: 
\begin{eqnarray}
\vec \tau_{\mathrm{grav}}=\frac{GM}{\xi^2}\left[m(\hat{\vec \xi}+ \vec \rho_\mathrm{grav})+ \frac{3}{\xi}(\hat{\vec \xi}\times(I\hat{\vec \xi}))\right]
\end{eqnarray}
where $\hat{\vec\xi}=T\cdot \hat{\vec x}$ is the position vector of the object transformed to the body system by the inertial to body transformation matrix $T$, $G$ is the gravitational constant, $M_{\Earth}$ the mass of the Earth, $m=\sum_k m_{k}$ the total mass of the object, $\vec \rho_\mathrm{grav}$ is the distance of the geometric center of the object to the center of mass.  \\
\\
The non-gravitational perturbations are modeled for the orbital as well as the attitude motion as represented by the last term of equation \ref{11}.  Direct radiation pressure is by far the largest perturbation in the geosynchronous orbital region. Radiation pressure perturbation directly depends on the current area and its orientation exposed to the radiation source. The body shape is approximated as sum over $n$ flat sub-area facets, the perturbing radiation force can be expressed at the following:
\begin{eqnarray}
\vec F_{\mathrm{rad,i}}=\textgoth{A}_i\frac{E}{c}
\frac{A_{\Earth}^2}{|\vec x-\vec x_{\Sun}|^2}
\hat{\vec S}\hat{\vec N_i} [(1-C_{\mathrm{s},i})\hat{\vec S}+2(C_{\mathrm{s},i}\cdot \hat{\vec S}\hat{\vec N_i}+\frac{1}{3}C_{\mathrm{d},i})\hat{\vec N_i}]\\
\mbox{for:}\hspace{0.5cm}0<\arccos( \hat{\vec S}\hat{\vec N_i})<\pi/2\hspace{0.5cm}\mbox{with:}\hspace{0.5cm}C_{\mathrm{s},i}+C_{\mathrm{d},i}=1-C_{\mathrm{a},i}\nonumber
\label{rad}
\end{eqnarray}
$m$ is the total mass of the satellite, $E$ is the solar flux, $A_{\Earth}$ the astronomical unit, $\vec x_{\Sun}$ the geocentric position of the sun, $c$ velocity of light, $\vec S$ the direction of the radiation source, $\textgoth{A}_i$ is the area of the i-th sub-area and $\vec N_i$ the normal vector of it. Back and front side are simulated at two different facets with two different normal vectors.  $C_{\mathrm{s,d,a},i}$ are the coefficients for specular, diffuse reflection and absorption for the i-th sub-area facet. The radiation over each sub-area is assumed to be constant, or to say in other words, the facets need to be chosen small enough to satisfy this condition at every point in time at the required confidence level. The direct radiation torque $\vec \tau_\mathrm{rad}$ as attitude perturbation and the acceleration $\vec a_{\mathrm{rad}}$ as orbital perturbation are determined as the following:
\begin{eqnarray}
\vec \tau_\mathrm{rad}=-\sum_{i=1}^n ( \vec \rho_\mathrm{grav}+\vec\rho_{\mathrm{rad},i})\times (T\cdot \vec F_{\mathrm{rad},i})
 \hspace{1cm}\vec a_{\mathrm{rad}}=\frac{\sum_{i=1}^n \vec F_{\mathrm{rad,i}}}{m}
\label{aandn}
\end{eqnarray}
$\vec\rho_\mathrm{rad}$ is the vector from the geometric center of the object to the center of pressure, $ \bf{F}_\mathrm{rad}$ is the radiation force, defined in Eq.\ref{rad}, $T$ the transformation matrix from the geocentric reference system in the body reference system. Since the body is modeled as flat sub-area surfaces the center of pressure of each sub-area facet is chosen as the center of the sub-area facet, which holds if all radiation is identical over the area over each of the the sub-facets. Self-shadowing would need to be modeled for shapes with concavities, the convex objects being addressed in this work do not suffer self-shadowing and therefore not require this for accurate modeling. 

\section{Earth's shadow}
Earth eclipse shadowing is modeled. Three different models have been implemented and are compared to each other. The first model, labeled $s1$ is the simplest possible geometric model, assuming a perfectly cylindrical shadow cone. It is assumed that Earth shadow is entered, if the angle $\beta$ between the geocentric unit vector to the sun and the orbital plane of the eclipse is smaller than:
\begin{eqnarray}
\beta<\arcsin\left(\frac{a_{\Earth}}{a}\right),
\label{cyl}
\end{eqnarray}
where $a_{\Earth}$ is the mean Earth radius and $a$ the semi-major axis of a circular
satellite orbit. This shadow cylinder is derived under the assumption of a spherical Earth. The boundary between the sunlight and eclipsed part of space is
assumed to be cylindrical, i.e., no distinction is made between umbra and penumbra. The Earth's atmosphere is neglected. Although this is a very simple model it has been proven useful in many applications with low area-to-mass ratio objects. A more advanced geometrical model, labeled, $s2$, is determining the actual fraction of the visible sun surface ($F_p$) at the object's position: 
\begin{eqnarray}
\tilde{E}=F_p\cdot E,
\label{F_p}
\end{eqnarray}
where $E$ is the incident total solar flux when the whole sun disc is visible, that is the solar constant. The model assumes an extended sun, but ignores limb darkening, which leads to the following set of equations, named Baker functions \cite{baker1967}:
\begin{eqnarray}
&&F_p=1-\frac{(\gamma/\tau)^2[(\delta-\sin(2\delta)/2)+(\beta-\sin(2\beta)/2)]}{\pi}\\
&\mbox{with}&\nonumber\\
&&\gamma=\arcsin(a_{\Earth}/x) \hspace{1cm} \tau=\arcsin(a_{\Sun}/x_{\mathrm{sat}\Sun})\nonumber\\
&&\epsilon=\arccos(\hat{\vec x}_{\mathrm{sat}\Sun}\hat{\vec x}) \left\{\begin{array}{cl} &\epsilon > \gamma+\tau \hspace{0.8cm}\mbox{whole sun disc visible}\\\nonumber
&\tau+\epsilon<\gamma \hspace{0.8cm}\mbox{no sun disc is visible}\\ &\mbox{otherwise} \hspace{0.8cm} \mbox{fraction of the sun disc visible}\end{array}\right.\\\nonumber
&&s=(\tau+\gamma+\epsilon)/2  \hspace{1cm} k=\sqrt{(s(s-\tau)(s-\gamma)(s-\epsilon))}\\\nonumber
&&\sin(\delta)=\frac{2k}{\epsilon\gamma} \hspace{1cm} \sin(\beta)=\frac{2k}{\epsilon\tau} 
\\\nonumber
&&\cos(\delta)=\frac{\epsilon^2+\gamma^2-\tau^2}{2\epsilon\gamma}\hspace{1cm}  \cos(\beta)=\frac{\epsilon^2+\gamma^2-\gamma^2}{2\epsilon\tau}
\label{F_p_1}
\end{eqnarray}
whereas $a_{\Sun}$ is the radius of the sun, $a_{\Earth}$ the mean Earth radius, $\hat{\vec x}_{\mathrm{sat}\Sun}$ is the unit vector from the satellites to the sun, with $| \vec x_{\mathrm{sat}\Sun}|=x_{\mathrm{sat}\Sun}$ the distance between satellite and sun, $\hat{\vec x}$ is the unit vector from the Earth to the satellite, and $|\vec x|=x$ the distance, respectively. Suggestions for increasing Earth or Sun radius have not been adopted since the comparison with more physical models do not show significant improvements \cite{hujsak_shadow,link_shadow}. All geometrical models have significant deficiencies. The most accurate shadow models are based on exact modeling of atmosphere and based on ray tracing techniques \cite{Vokrou1,Vokrou2}. But they impose a tremendous computational burden, which is seeked to be prevented in the six dimensional orbit attitude integration. Alternatively, the dual-conic model may be fit to the more physical model in a weighted sum of the Baker functions according to Hujsak \cite{hujsak_shadow}:
\begin{eqnarray}
&&\tilde{F}_p=\sum_i^5 w_iF_p(\tilde{x}_i)\hspace{1cm}\mbox{with}\\
&&\{\tilde{x}_i\}=\{a_{\Earth}+48\,\mathrm{km},a_{\Earth}+32\,\mathrm{km},a_{\Earth}+16\,\mathrm{km},a_{\Earth},a_{\Earth}-40\,\mathrm{km}\},\\\nonumber
&&w_i=\{2/9,2/9,2/9,2/9,1/9\},
\label{F_p_2}
\end{eqnarray}
where $a_{\Earth}$ is the mean Earth radius, $w_i$ are the weighting parameters, and  $F_p$ is evaluated with Eq.\ref{F_p_1}, using the components of $\{\tilde{x}_i\}$ instead of the Earth radius.

\section{Light Curves}
The solar radiation not only exercises a force, but also illuminates the object. The reflected radiation, which may be observed by a ground based sensor, is thus a function of the observation geometry and distance from the observer, as well as the shape, size, material properties and orientation of the object surfaces. The integrated signal is measured as a so-called light curves, that is measurement of the brightness variations over time. The magnitude measured by a ground based optical sensor is determined as the following:
\begin{eqnarray}
\mathrm{mag}=\mathrm{mag}_\Sun-2.5\log \textgoth{G}(\vec x)
\end{eqnarray}
For a sphere of radius $r$ the reflection function \textgoth{G} is determined as:
\begin{eqnarray}
\textgoth{G}(\vec x)=\frac{4\cdot r^2 \cdot C_{\mathrm{d}}\cdot(\sin(\theta)+(\pi-\theta)\cos(\theta))}{(x-a_\Earth)^2},
\label{fluxball}
\end{eqnarray}
$\theta$ is the angle between the observation direction $\hat{\vec O}$, from the station to the object,  and the direction of the illumination source $\hat{\vec S}$. Specular reflection in the direction of the observer is not possible.\\
\\
For an object consisting of $n$ flat surfaces with areas $\textgoth{A}_i$ with the corresponding normal vectors $\hat{\vec N_i}$ the reflection function reads as:
\begin{eqnarray}
\textgoth{G}(\vec x)=\left(\sum_{i=1}^n\frac{\textgoth{A}_i}{\pi\cdot(x-a_\Earth)^2}\big [C_{\mathrm{d},i}\hat{\vec O}\hat{\vec N_i}\cdot \hat{\vec S}\hat{\vec N_i}+\frac{\tau_i\cdot C_{\mathrm{s},i}\cdot x_\Sun^2}{a_\Sun^2}\big ]\right)\\
\tau_i=\left\{\begin{array}{clll} 
1  &\mbox{for }1- \cos(0.25\deg)\leq\frac{\vec O+\vec S}{|\vec O+\vec S|}\cdot \hat{\vec N_i}\leq 1+ \cos(0.25\deg)\\
0  &\mbox{else}
\end{array}\right.\nonumber
\label{fluxplate}
\end{eqnarray}
for $0<\arccos( \hat{\vec S}\hat{\vec N_i})<\pi/2$ and $0<\arccos( \hat{\vec S}\hat{\vec O})<\pi/2$. $\mathrm{mag}_\Sun$ is the magnitude of the sun, $\vec O$ the viewing direction of the sensor, $\tau_i$ the specular reflection parameter, determining if specular reflection condition is met. The dimension of the sun has been taken into account, where $ a_\Sun$ is the radius of the Sun, $ a_\Earth$ the mean radius of the Earth.  No limb darkening effects have been accounted for. 

\section{Sensor Model}
If the object is visible in principle and within the field of view, the so-called signal to noise ratio (SNR) is the measure of the detection probability and is a measure for the magnitude uncertainty of the detection. Combining the models explored in \cite{dissholger,merlinehowell}, the SNR can be defined as given in Eq.\ref{SNR}.
\begin{eqnarray}
&&\mathrm{SNR}=\\&&\frac{S_{\mathrm{obj}}}{\sqrt{S_{\mathrm{obj}}+n_{\mathrm{pix}}(S_{\mathrm{stars}}+S_{\mathrm{gal}}+S_{\mathrm{zodi}}+S_{\mathrm{airgl}}+S_{\mathrm{atscat}}+\mathrm{d}\cdot t_{\mathrm{int}}+\mathrm{r}^2+\mathrm{g}^2\sigma_g^2})}.\nonumber
\label{SNR}
\end{eqnarray}
The signal terms in units of electrons are the signal of the object itself ($S_{obj}$), and signals of the different background noise sources: the stars ($S_{stars}$), galaxies ($S_{gal}$), zodiac light ($S_{zodi}$), airglow ($S_{airglow}$), atmospheric scattering ($S_{atscat}$), as well as direct contribution of CCD readout $r$ and dark noise $d$, and the gain factor $g$ with its standard deviation $\sigma_g$. Full linearity of the gain is assumed. $n_{\mathrm{pix}}$ is the number of pixels, over which the object signal is spread. The flux of the object is determined according to Eq.\ref{fluxball} and Eq.\ref{fluxplate}, respectively. The flux per pixel of the object is determined as the following:
\begin{eqnarray}
\tilde{\textgoth{G}}(\vec x)=\frac{\textgoth{G}(\vec x)}{l(t_{\mathrm{int}})}\hspace{0.5cm}\mbox{with}\hspace{0.5cm}
l=\frac{|\vec v_{\mathrm{||tel}}-\vec v_{\mathrm{||obj}}|\cdot t_{\mathrm{int}}}{\mathrm{scale}}
\end{eqnarray}
 $l(t_{\mathrm{int}})$ is the so-called pixel dwell time function, $\vec v_{\mathrm{||tel}}$ is the velocity of the telescope pointing during the exposure, and $\vec v_{\mathrm{||obj}}$ the object's velocity during the exposure parallel to the image plane, $scale$ denotes the pixel scale, and $ t_{int}$ is the integration time.\\
\\
The count rates of the different sources are determined via the individual fluxes, which we averaged over the complete visible band pass, see Eq. \ref{count} \cite{dissholger}: 
\begin{eqnarray}
C(\vec x)=\frac{\pi}{4}\cdot\mathrm{D}\cdot \tilde{\textgoth{G}}(\vec x)\cdot\frac{\bar\lambda}{hc}\cdot\mathrm{QE(\bar\lambda)}e^{(\kappa(\bar\lambda)\cdot \sec(\phi_{\mathrm{zenith}}))}\cdot \Delta\lambda
\label{count}
\end{eqnarray}
$\vec x$ is the position of the object, $D$ is the aperature of the telescope, $\tilde{\textgoth{G}}(\vec x)$ the flux per pixel rate of the object and the background sources, $c$ the speed of light, $h$ Planck's contant, $\bar\lambda$ is the wavelength bandwidths, we are averaging over, $QE$ is the quantum efficiency of the sensor, $\kappa$ is the atmospheric extinction coefficient, and $\phi_{\mathrm{zenith}}$ the zenith angle, which determines the amount of air mass the signal is passing through. The signal is defined as the count rate during the integration time $t_{\mathrm{int}}$:
\begin{eqnarray}
S(\vec x_{\mathrm{obj}}(t))=C(\vec x)\cdot t_{\mathrm{int}}.\label{signal}
\end{eqnarray}
The individual fluxes per pixel of the background sources are determined according to Krag\cite{dissholger}, the measured fluxes values and the extinction coefficient are taken from Daniels\cite{73}, Allen\cite{star_allen}. For the stellar background, no exact positions of any of the star catalogues are read in, so no eclipsing with the observed object is determined. Instead, only the fainter stars are taken into account and are treated as continuous background sources and their common flux is smeared over the field of view (FOV). The Guide Star Catalog (GSC) catalog values for star densities have been patched towards higher magnitudes \cite{dissholger}. Scattered moonlight is not considered here, but no detectability of any object signal is assumed in the halo of the moon. \\
The sensor model allows to determine the actual standard variation in the magnitude in the simulated light curves, which are the following \cite{merlinehowell}:
\begin{eqnarray}
\sigma_\mathrm{mag}=1.0857\cdot \frac{1}{\mathrm{SNR}}
\label{magi}
\end{eqnarray}
\begin{figure}[h]
  \centering
  \subfloat[]{\includegraphics[width=0.55\textwidth]{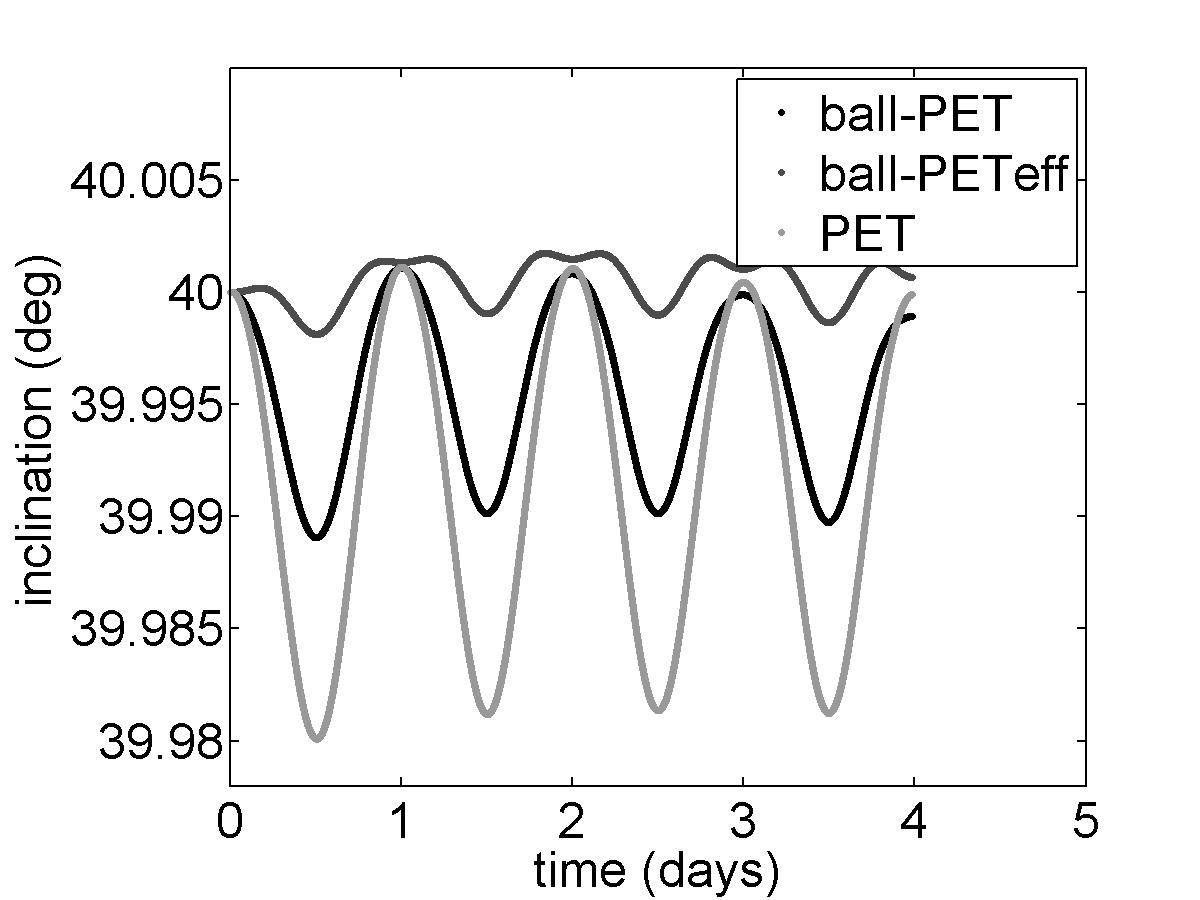}} 
  \subfloat[]{\includegraphics[width=0.55\textwidth]{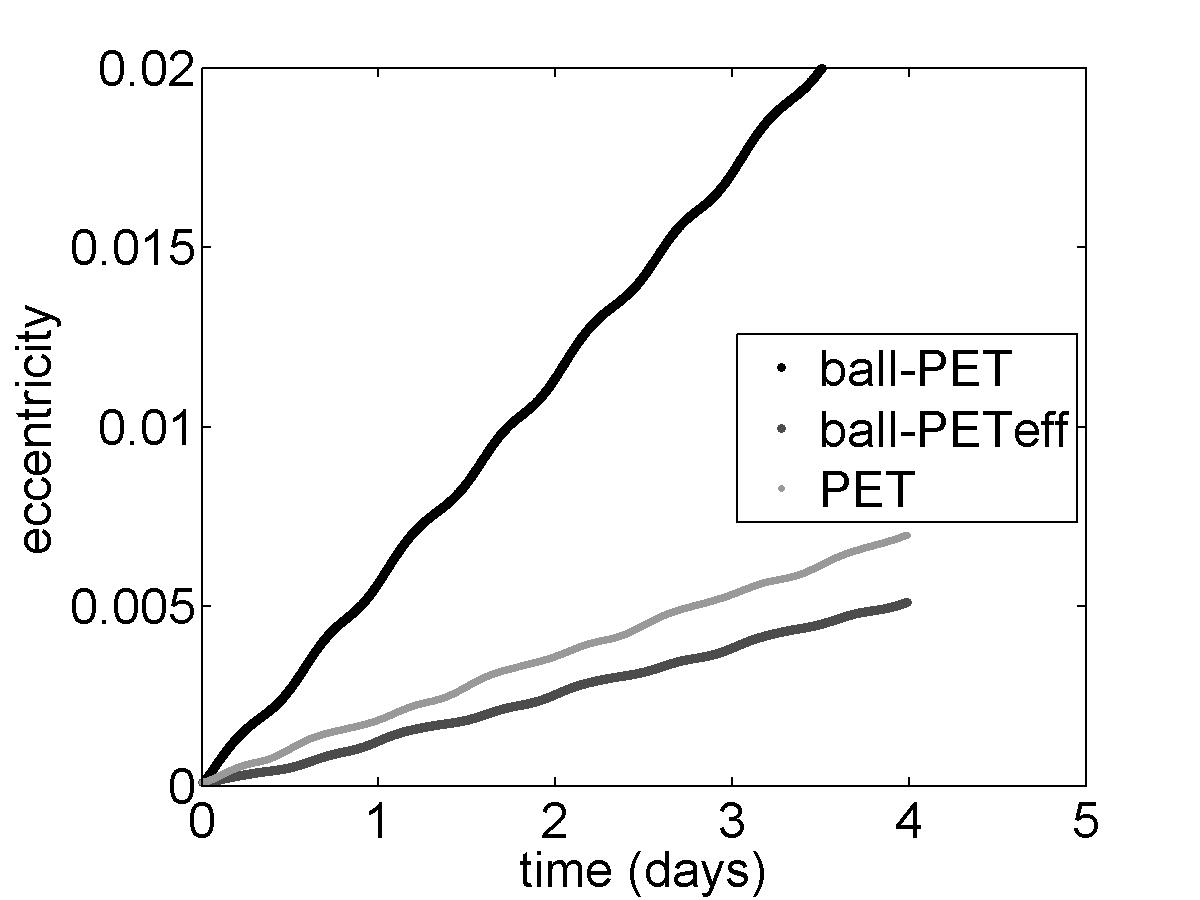}}
  \caption{\bf Inclination and eccentricity evolution for \textit{ball-PET},  \textit{ball-PETeff},  \textit{PET}.}
  \label{pet0incecc}
\end{figure}
\begin{figure}[h!]
  \centering
  \subfloat[]{\includegraphics[width=0.55\textwidth]{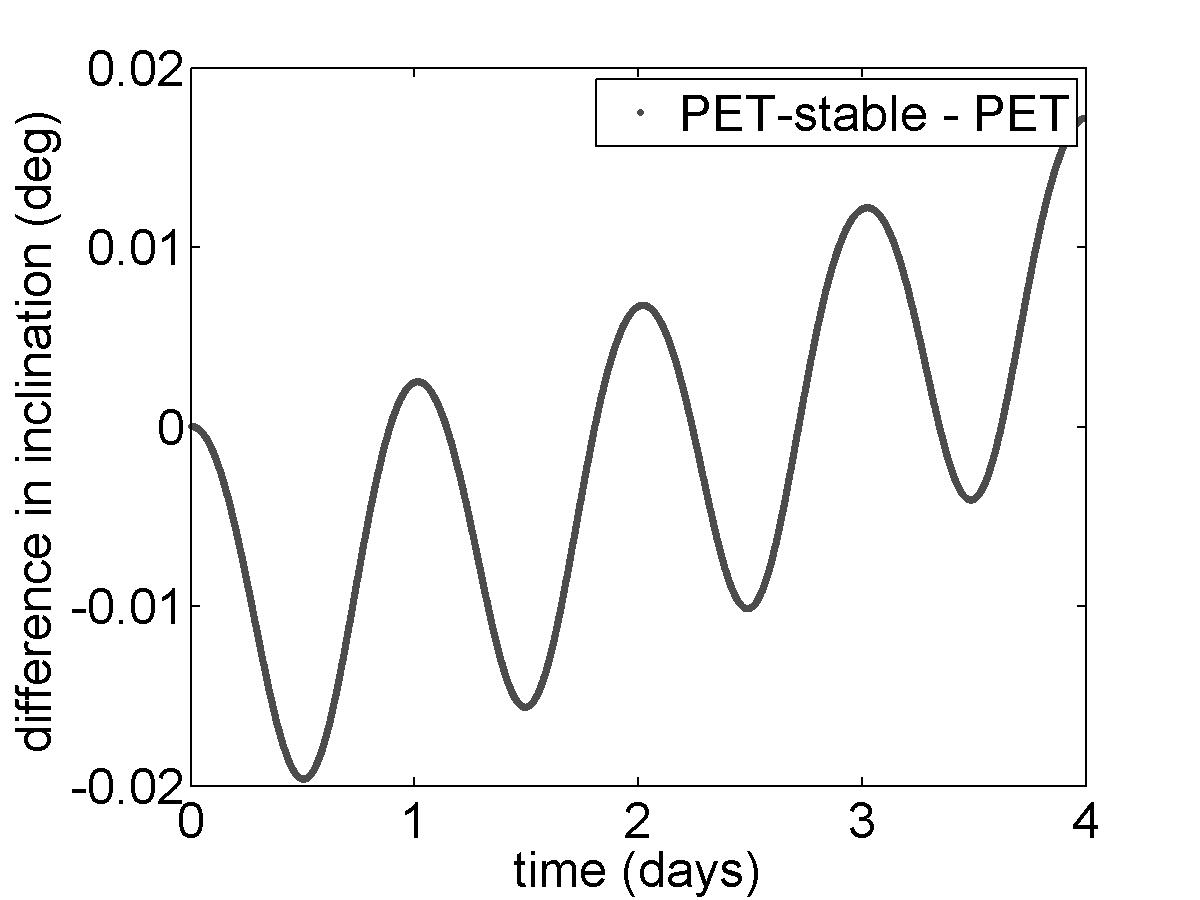}} 
  \subfloat[]{\includegraphics[width=0.55\textwidth]{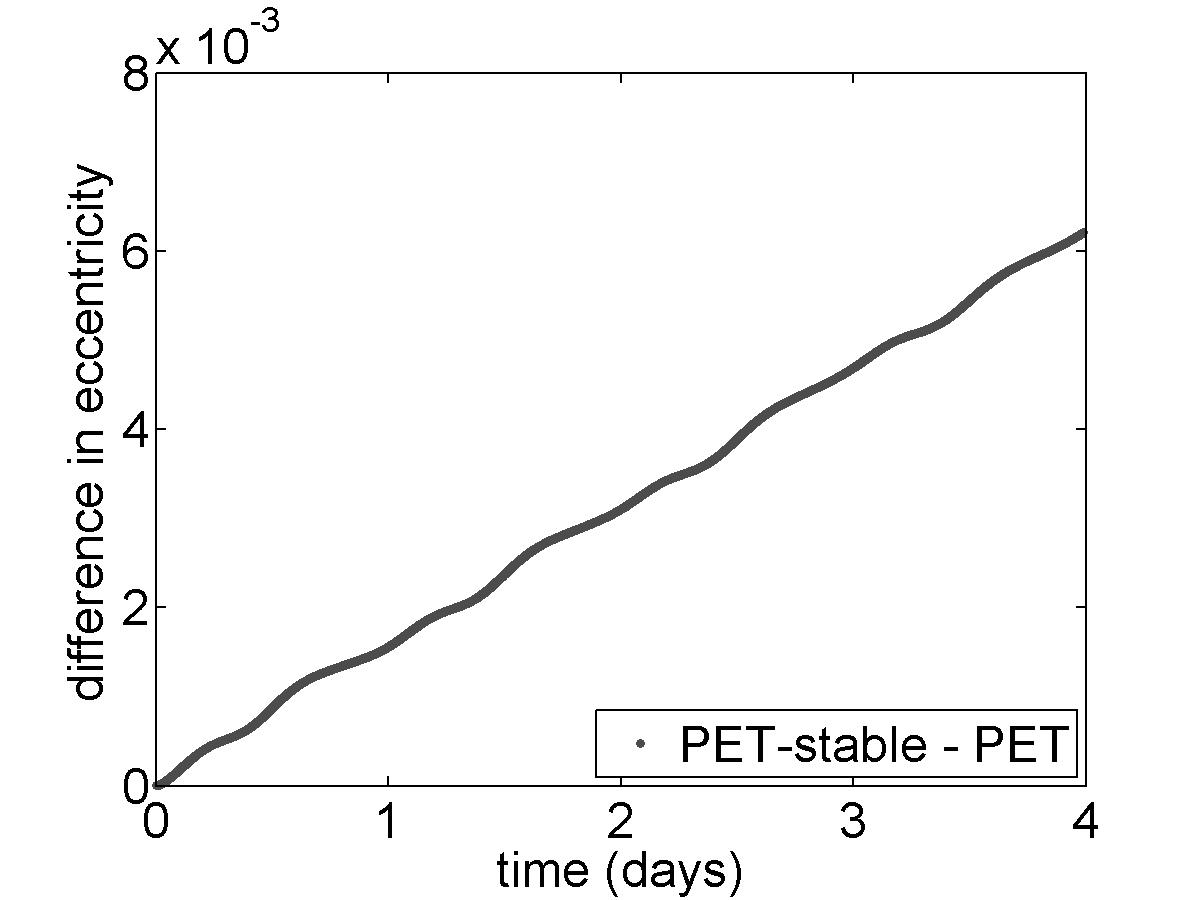}}
  \caption{\bf Difference in inclination and eccentricity between \textit{PET-stable} and \textit{PET}.}
  \label{pet0zoom}
\end{figure}

\section{Results}
The different object models have been run in several simulations. All objects start with the same set of initial osculating elements, which are a  initial semi-major axis of 42164.0$km$, initial mean anomaly is chosen to be -84.0 degrees,  argument of perigee of -351.0 degree and right ascension of ascending node of 60.0 degree, and a small eccentricity of 0.0001. Inclination is chosen to be 40.0 degrees. Anomaly is chosen so that the object might be observable in principle from a European observing station. High area-to-mass ratio objects tend are perturbed to reach high inclinations, even when originally stemming from zero inclination orbits, therefore a high inclination of 40 degrees has been chosen. Initial Euler angles are chosen to be 24.4, 53.8 and -15.0 degree (3-1-3 sequence), in the following the angles are referred to as first, second and third Euler angle. For the Euler angles an random orientation has been chosen, which does not align with the axis of th Earth or the direction of velocity. Although the calculation has been done in quaternions to avoid singularities, initial conditions and the results are presented in Euler angles for the sake of a clear physical interpretation. All objects start with no initial angular velocity.\\
\newpage
\begin{figure}[h!]
  \centering
  \subfloat[]{\includegraphics[width=0.55\textwidth]{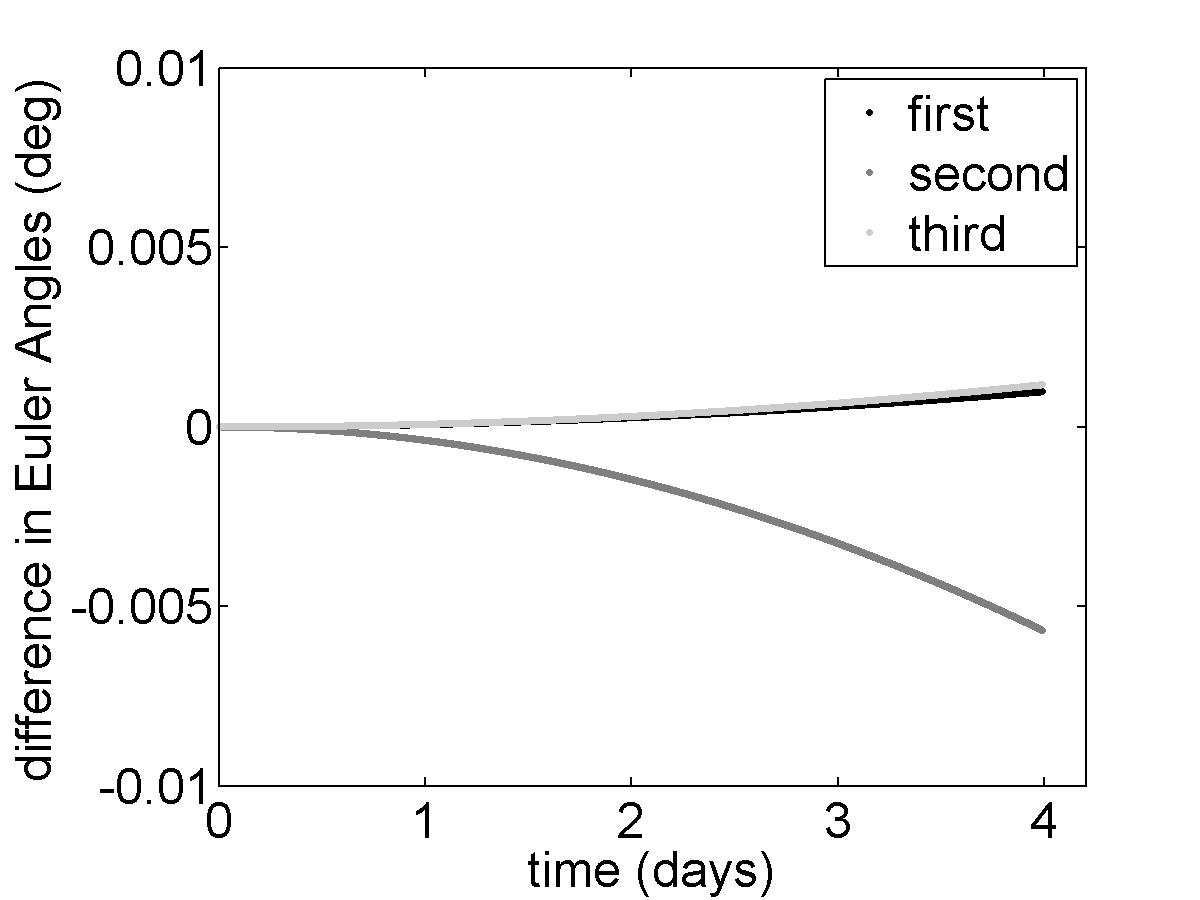}} 
 \subfloat[]{\includegraphics[width=0.55\textwidth]{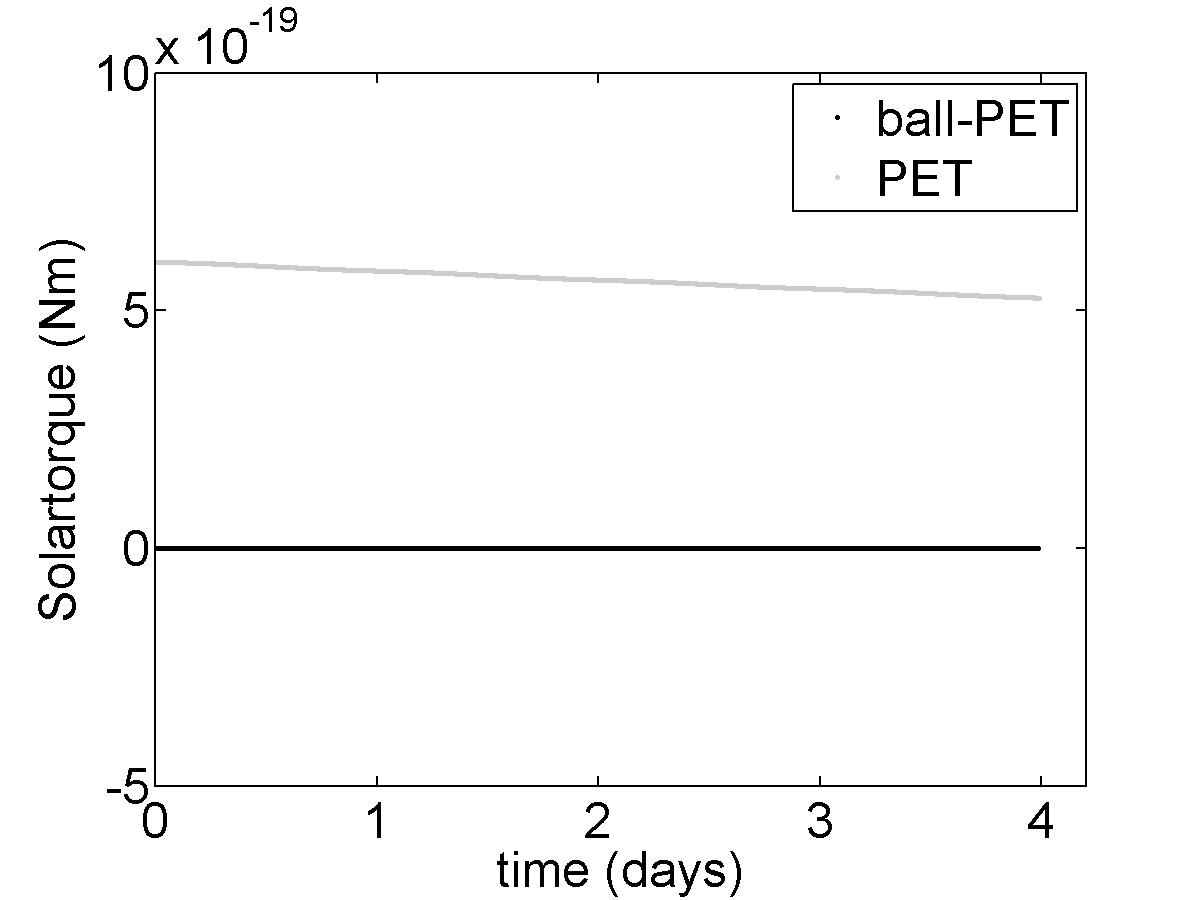}}
  \caption{\bf Evolution of the first, second and third Euler (3-1-3) angle and the direct radiation pressure torque of object \textit{PET}.}
  \label{pet0euler}
\end{figure}
\begin{figure}[h!]
  \centering
  \subfloat[]{\includegraphics[width=0.55\textwidth]{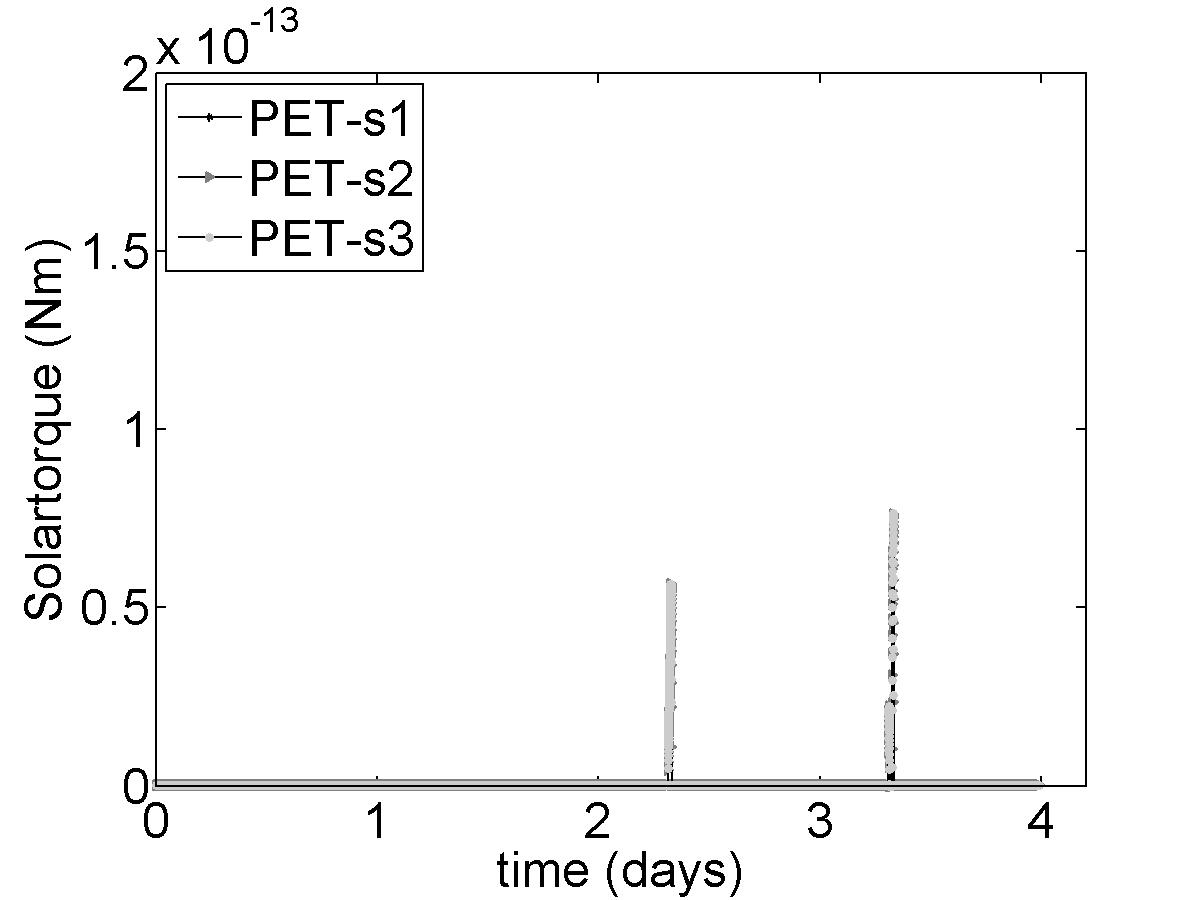}} 
  \subfloat[]{\includegraphics[width=0.55\textwidth]{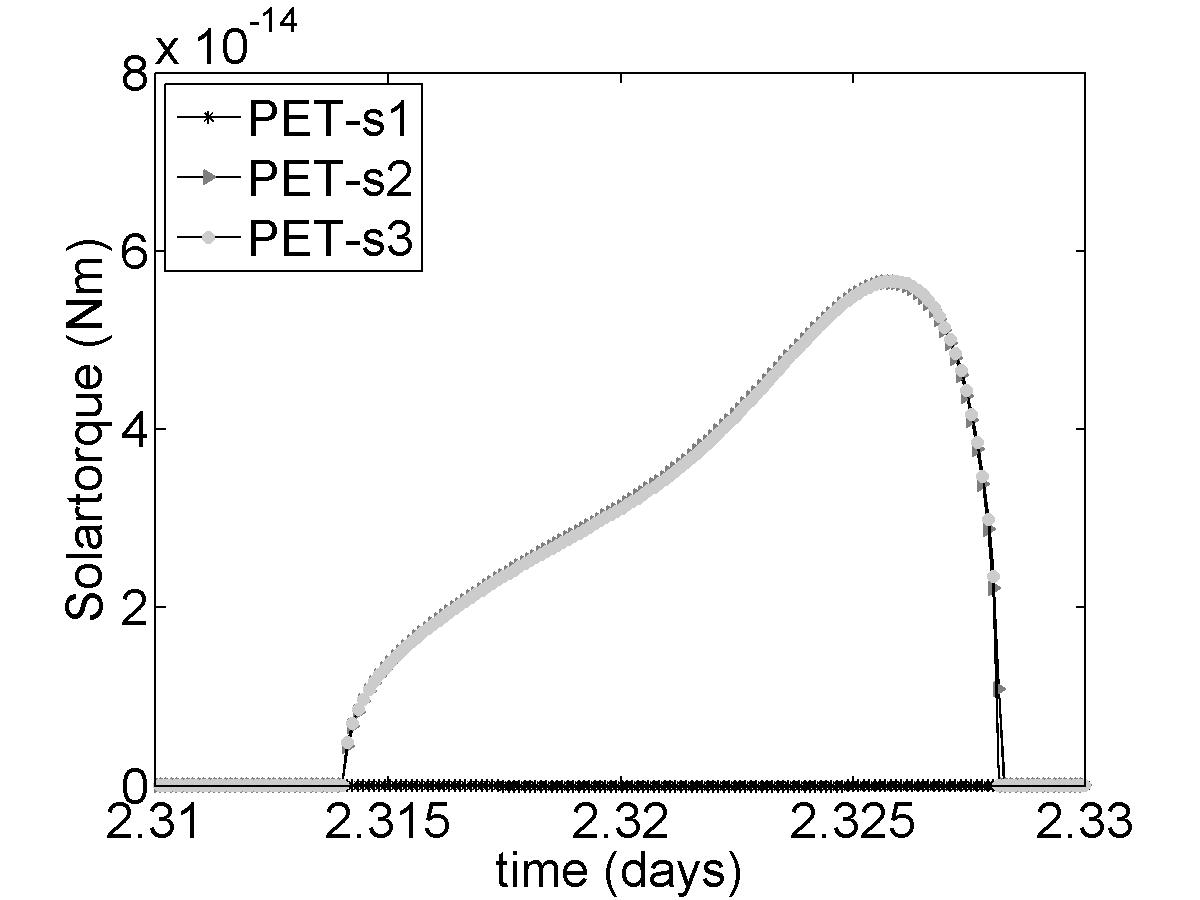}}
  \caption{\bf Solar torques at passages through Earth's shadow and for different shadow models: $s1$, $s2$, $s3$ (a) overview, (b) zoom on first Earth's shadow passage.}
  \label{torque}
\end{figure}
\newpage
\begin{figure}[h!]
  \centering
  \subfloat[]{\includegraphics[ width=0.55\textwidth]{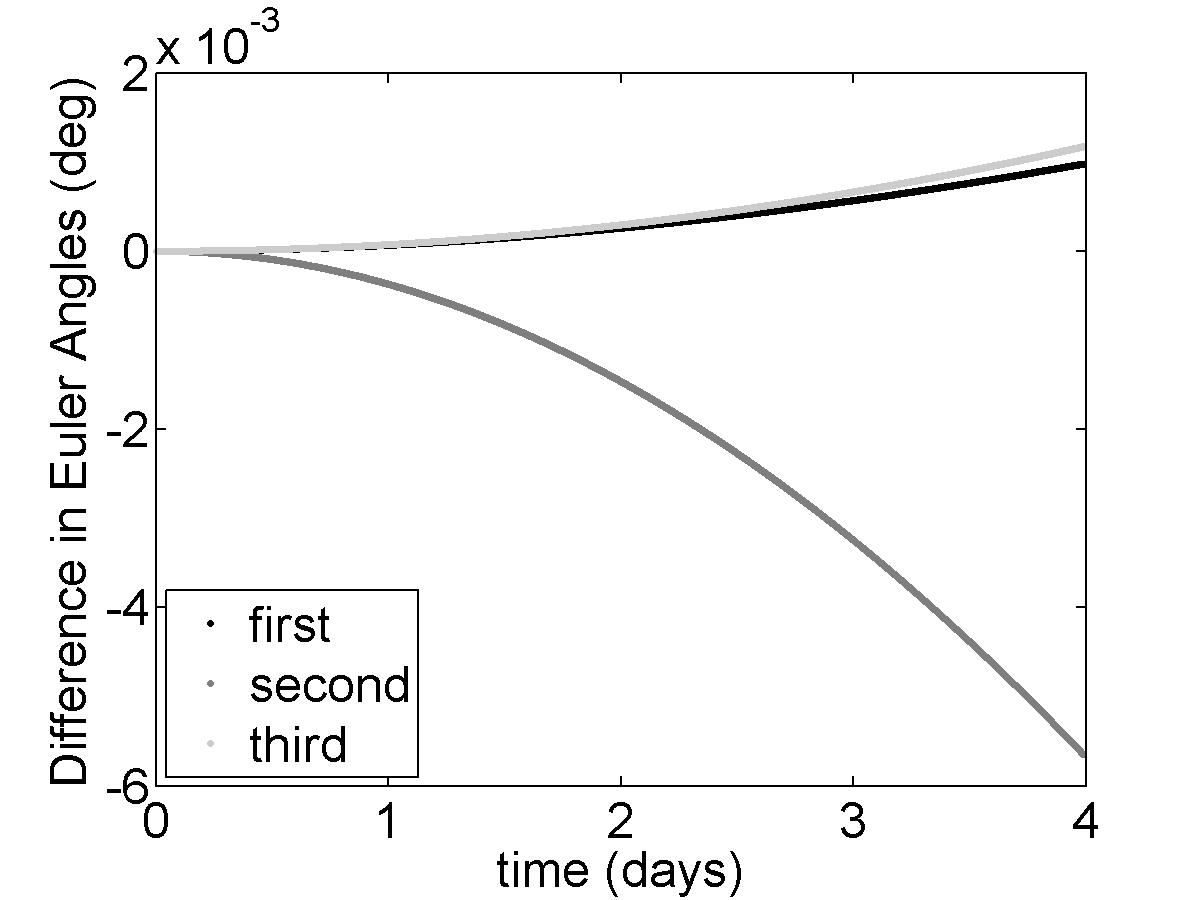}} 
  \subfloat[]{\includegraphics[ width=0.55\textwidth]{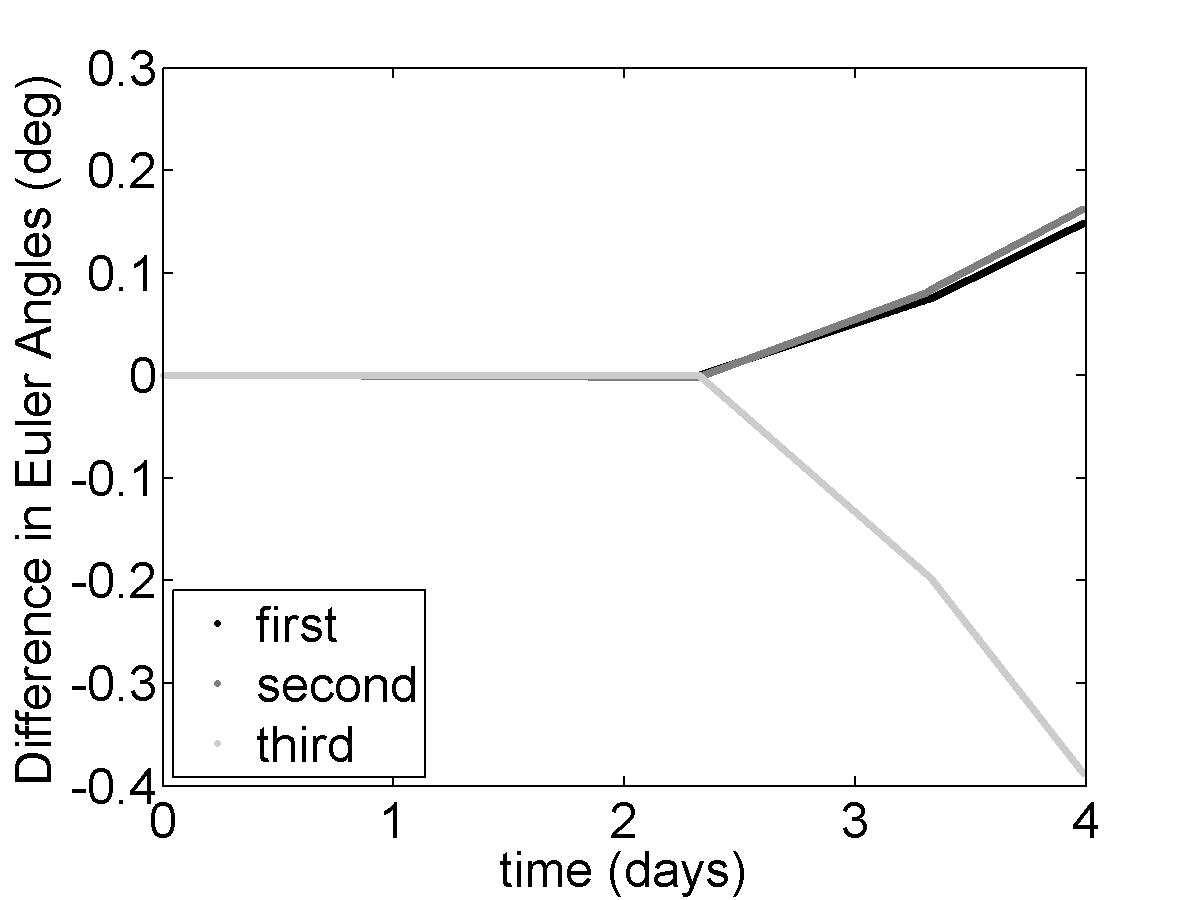}}\\
  \subfloat[\label{s2s3}]{\includegraphics[ width=0.55\textwidth]{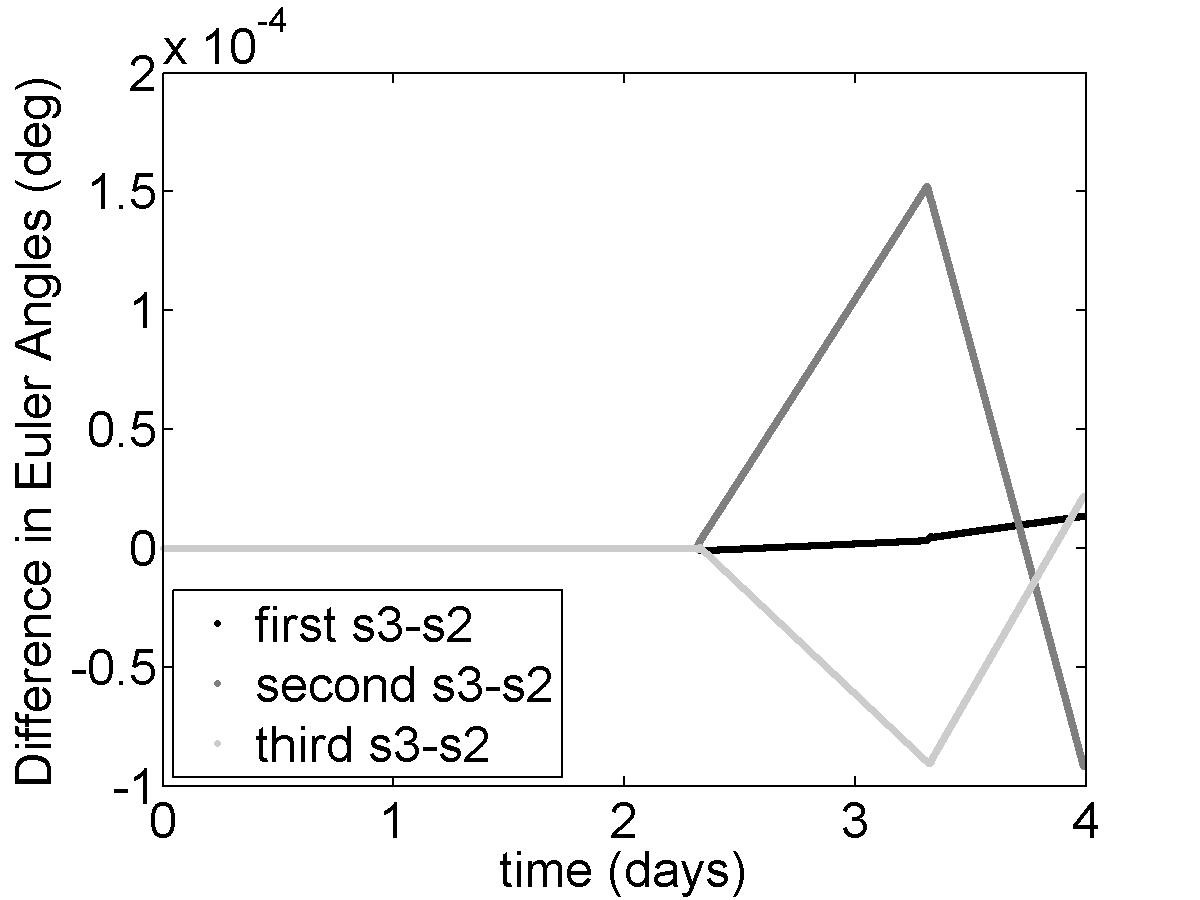}}
  \caption{\bf First, second and third Euler angle evolution of object \textit{PET} relative to the initial values for shadow model $s1$ (a), $s2$ (b), (c) difference between model $s3$-$s2$.}
  \label{euler_shadow}
\end{figure}
\begin{figure}[h!]
 \centering
  \subfloat[]{\includegraphics[width=0.55\textwidth]{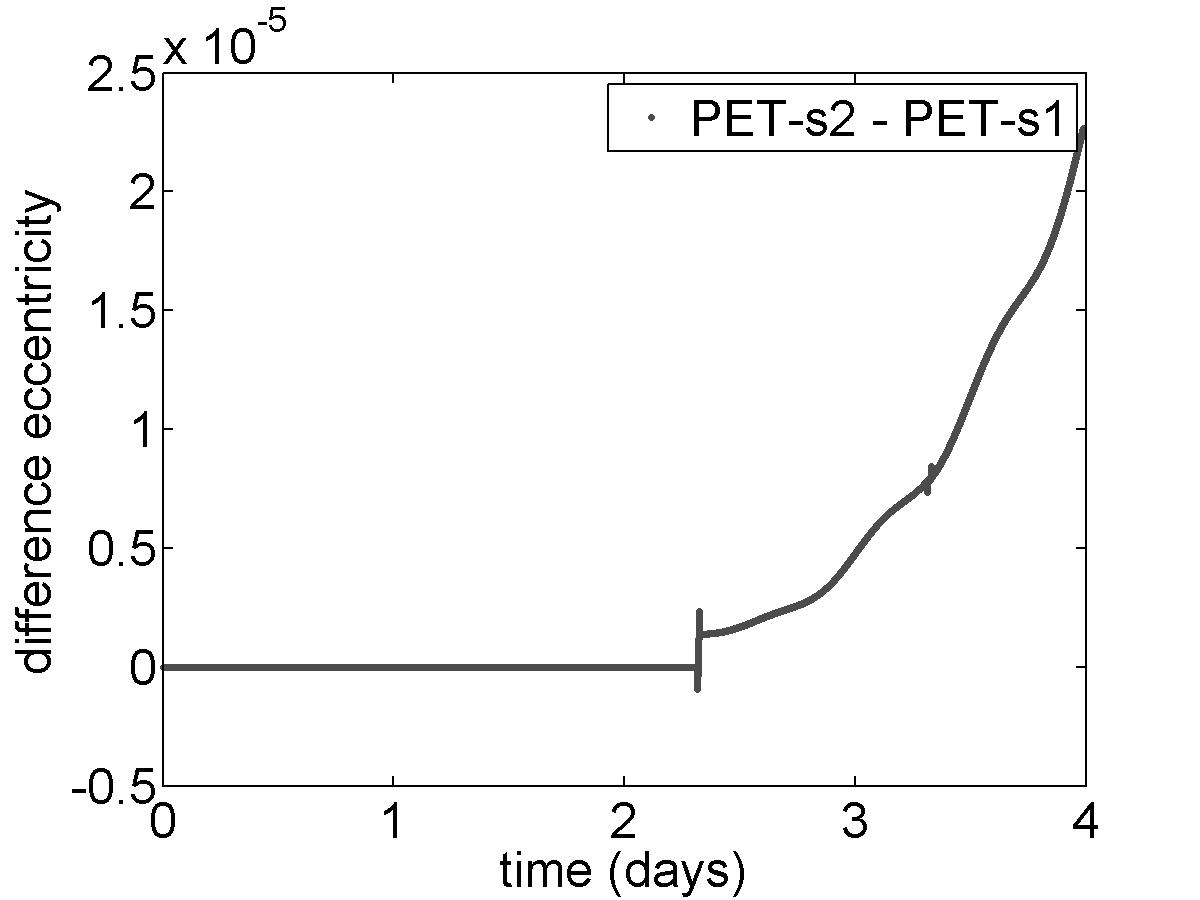}} 
  \subfloat[]{\includegraphics[width=0.55\textwidth]{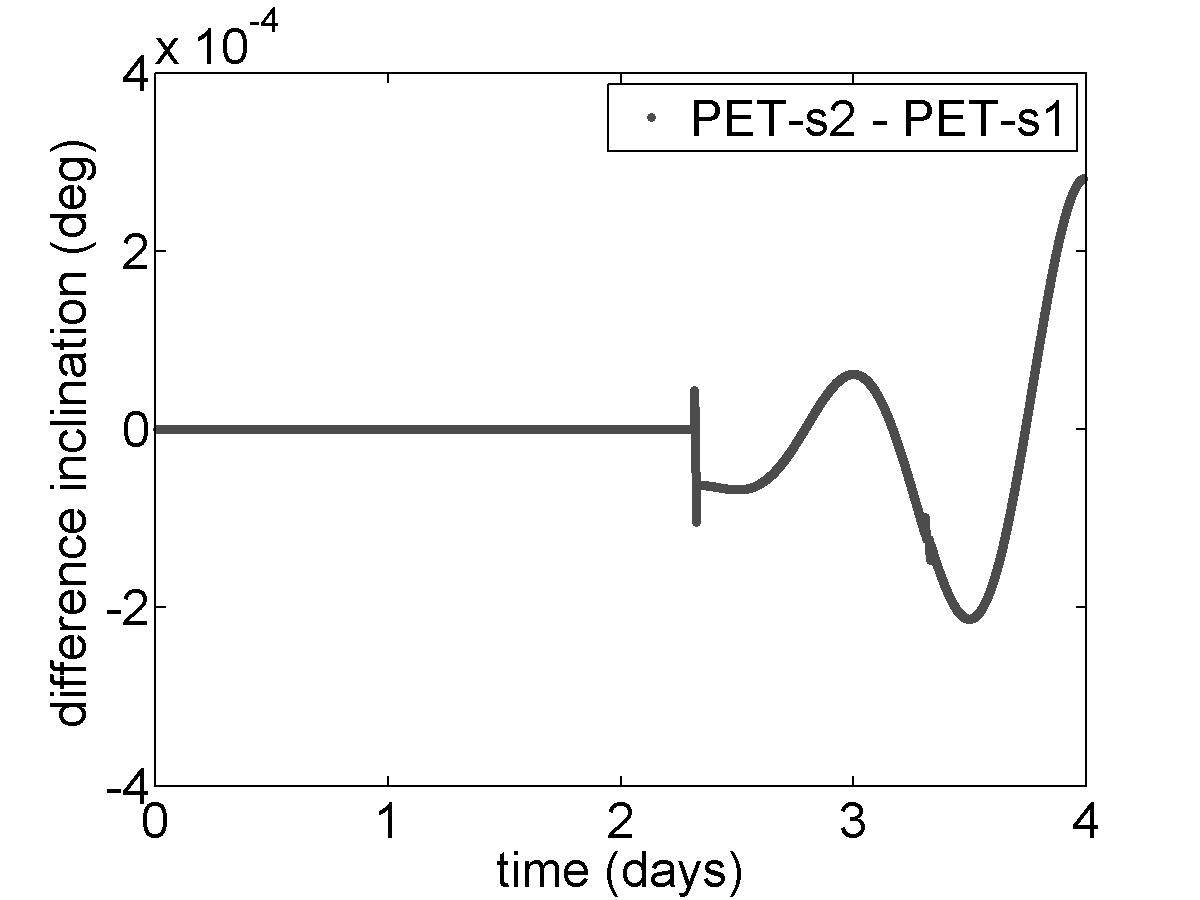}} 
 \caption{\bf Differences in the eccentricity and inclination evolution of object \textit{PET} between model $s1$ and model $s2$.}
 \label{eccinc_rel}
\end{figure}
Fig.\ref{pet0incecc} shows the comparison of the orbital elements inclination and eccentricity between the canon ball object \textit{ball-PET} and the object with the same physical AMR properties as the flat sheet object \textit{PET}, and the object \textit{ball-PETeff}, which exposes the same effective AMR at the start of the simulation than \textit{PET}. No Earth's shadow has been modeled.  Fig.\ref{pet0incecc} shows the periodic short term variations in inclination and eccentricity, as well as the secular trend towards higher eccentricities, which are typical for HAMR objects. Furthermore it reveals that the variations in the orbital elements are significantly different for both canon ball object, \textit{ball-PET} and \textit{ball-PETeff}, compared to the plate object \textit{PET}. The plate shaped objects show larger amplitudes in the variations in inclination, whereas in the increase in eccentricity it lies between the two spherical objects. The flat object exposes has a varying AMR over the course of one orbital revolution, it stays constant for the spherical objects. The total solar acceleration of the plate shaped object lies in between the two spherical objects, a figure is shown in the supplement materials. The chosen initial condition for the flat sheet lead to the fact that it is inclined to the line of radiation, therefore exposing overall a smaller net area than the spherical object, but in general also a larger one than the around 20 percent of the surfaces at the beginning of the simulation. to which \textit{ball-PETeff} is modeled. For the spherical object typical l daily variations in the acceleration due to direct radiation pressure caused by the varying distance to the Sun occur. A figure is shown in the supplement material.\\
\begin{figure}[h!]
  \centering
  \subfloat[]{\includegraphics[ width=0.55\textwidth]{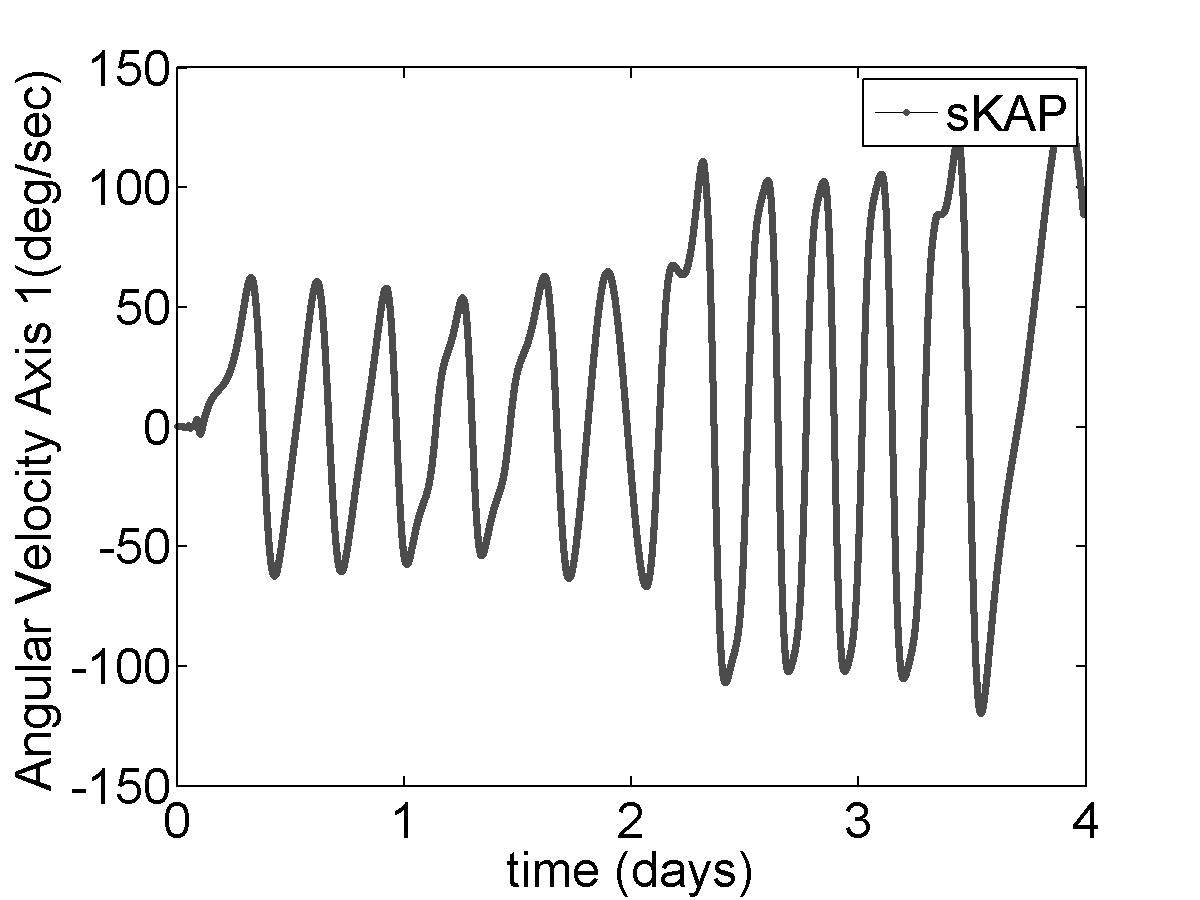}} 
  \subfloat[]{\includegraphics[ width=0.55\textwidth]{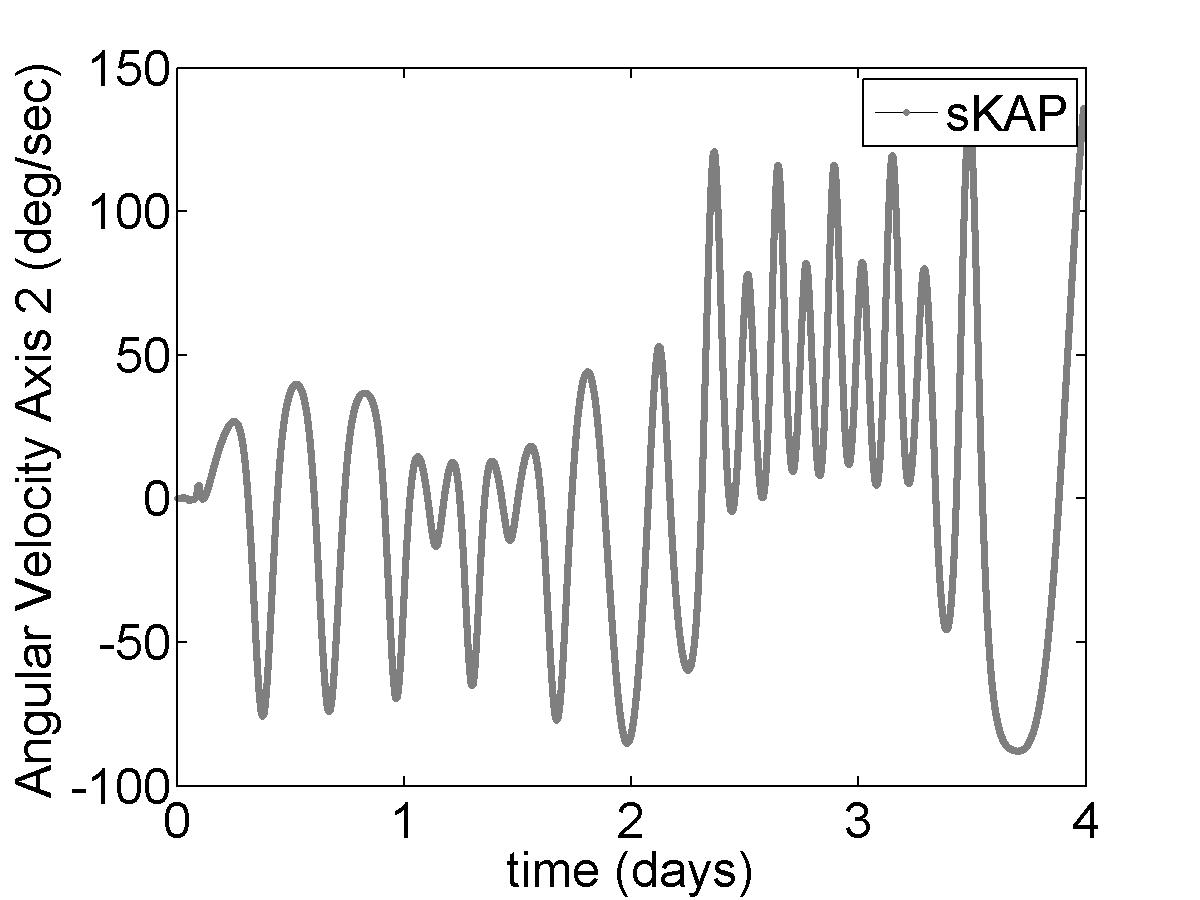}}\\
  \subfloat[]{\includegraphics[ width=0.55\textwidth]{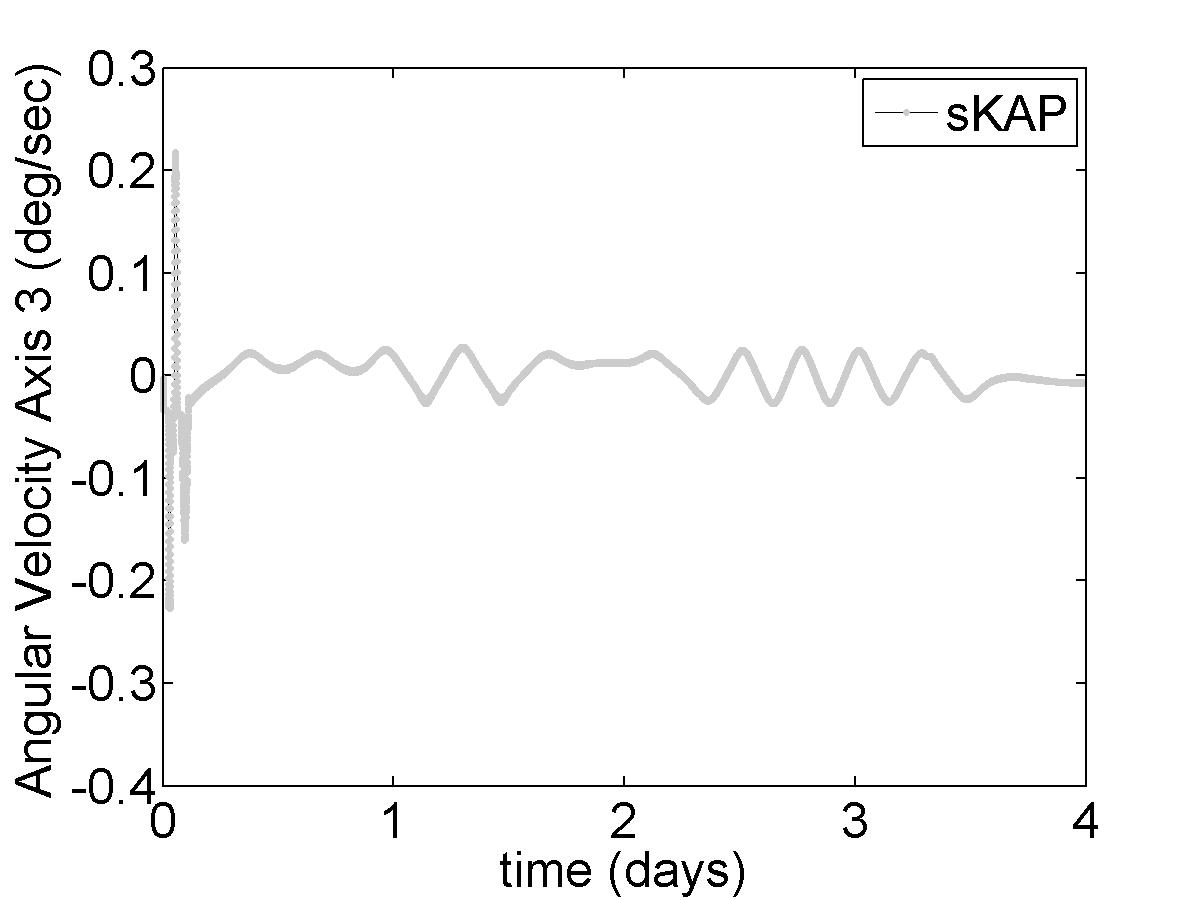}}
  \caption{\bf Angular velocity evolution of object \textit{sKAP}.}
  \label{sKAPomega}
\end{figure}
\begin{figure}[h!]
  \centering
  \subfloat[]{\includegraphics[ width=0.55\textwidth]{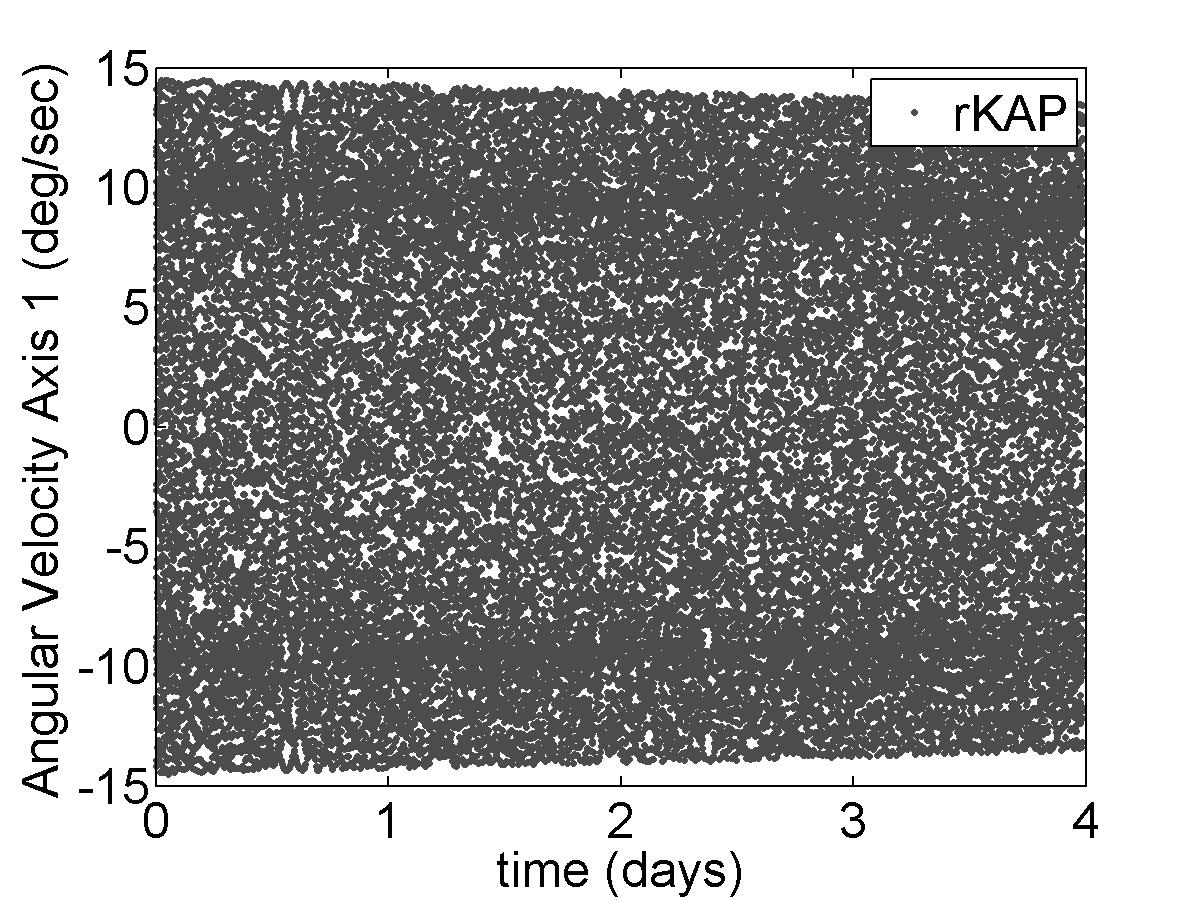}} 
  \subfloat[]{\includegraphics[ width=0.55\textwidth]{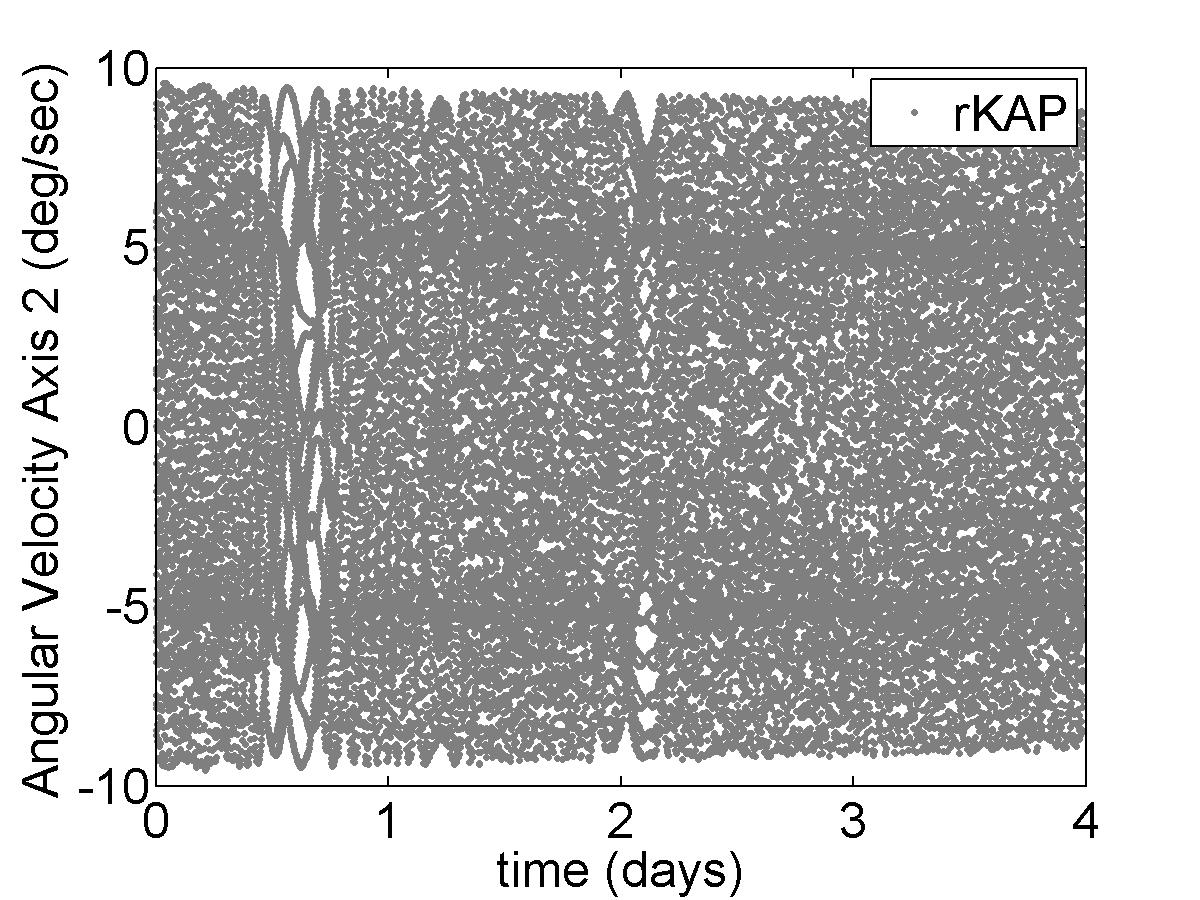}}\\
  \subfloat[]{\includegraphics[ width=0.55\textwidth]{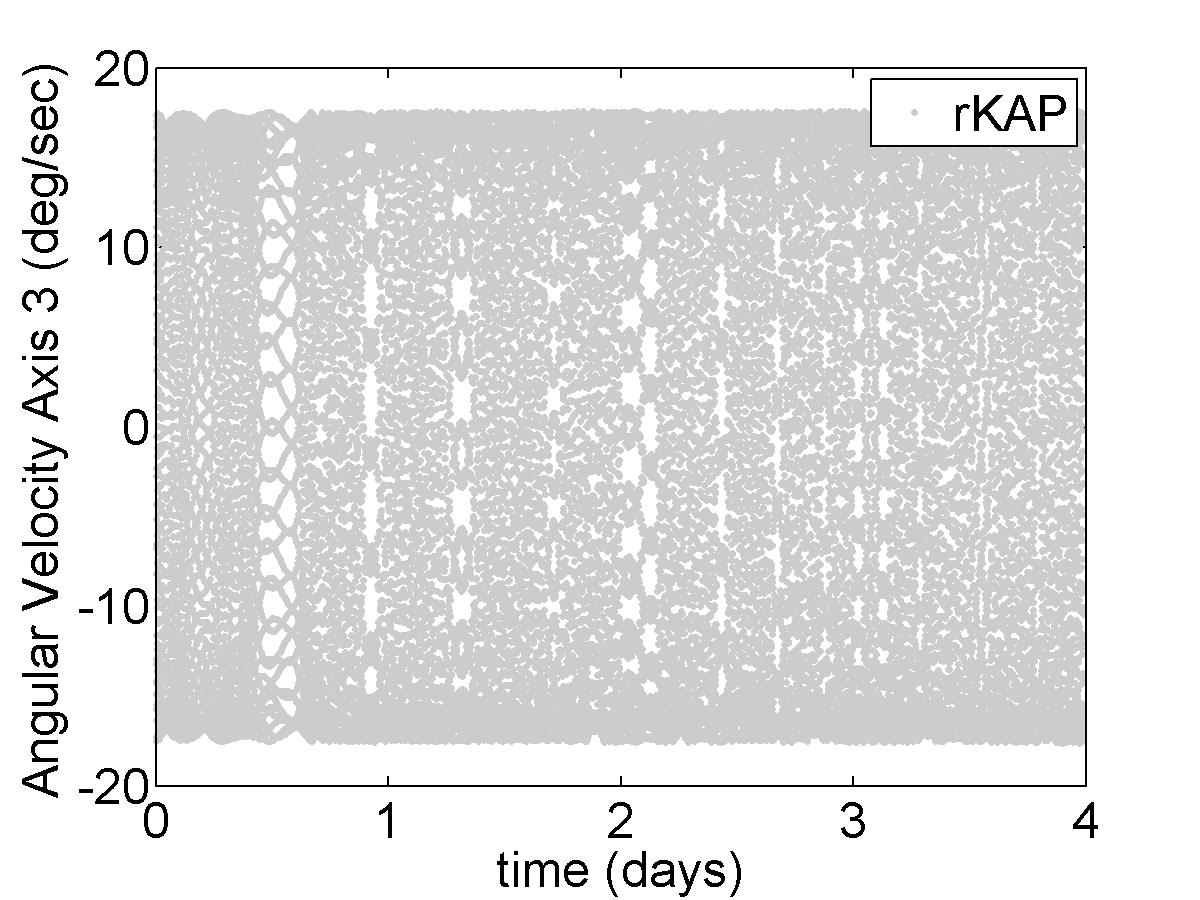}}
  \caption{\bf Angular velocity evolution of object \textit{rKAP}.}
  \label{rKAPomega}
\end{figure}
\begin{figure}[h!]
  \centering
  \subfloat[]{\includegraphics[ width=0.55\textwidth]{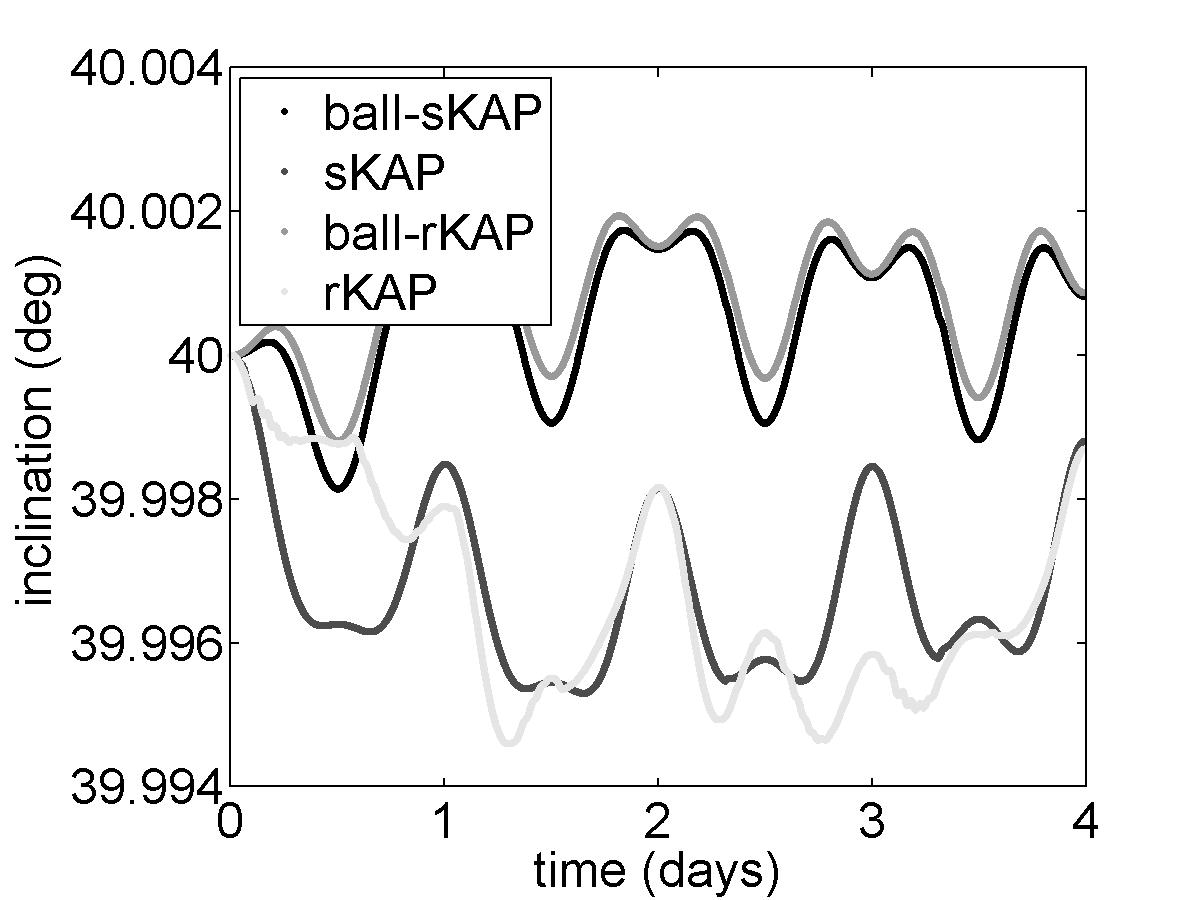}} 
  \subfloat[]{\includegraphics[ width=0.55\textwidth]{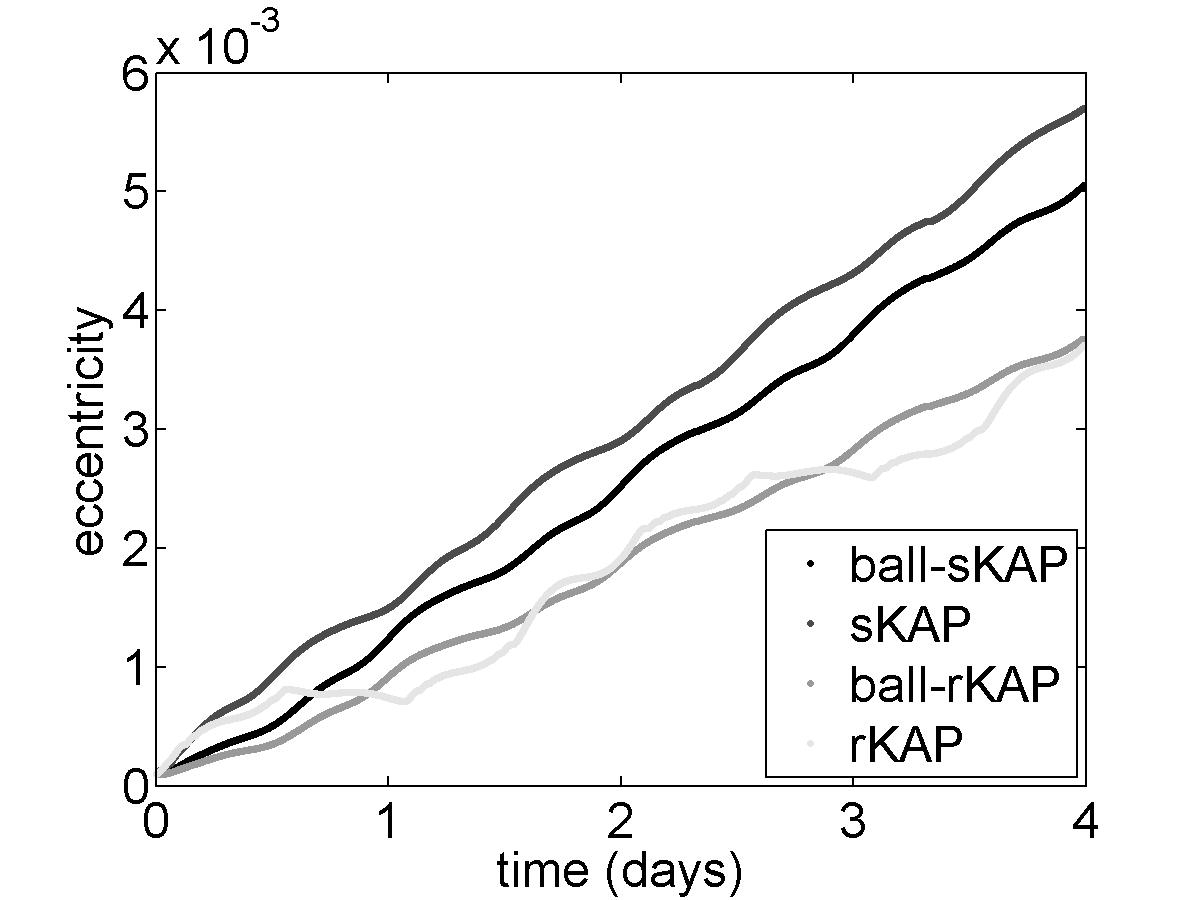}}
  \caption{\bf Inclination and eccentricity evolution of the two rigid plate objects \textit{sKAP} and \textit{rKAP}, in comparison to the spherical objects with the same reflection properties \textit{sKAP-ball} and \textit{rKAP-ball}.}
  \label{differ}
\end{figure}
\\
In a next step, the flat rigid-plate object is simulated twice: once under neglect of any attitude motion, labeled \textit{PET-stable} and once with a full six degree of freedom propagation, labeled \textit{PET}. Fig.\ref{pet0zoom} reveals that the orbital elements show difference of the order of $6\cdot 10^{-3}$ in eccentricity and of about $0.02$ degrees in inclination after the four day propagation interval.  Those very small differences arecaused by the fact that even though the object has perfectly uniform reflection properties, the attitude of the object slightly changes, because of a torque induced over the exposed area from one end to the other relative to the Sun. The change in Euler angles for object \textit{PET} and the induced solar radiation torque are shown in Fig.\ref{pet0euler}. Although these values are small in the investigated four day period, this secular effect may lead to significant errors in long term propagation of HAMR objects. \\
\\
The same objects are investigated again, this time including the modeling of shadow paths, which have been neglected so far. Three different shadow models have been used.  For the shadow model one, labeled $s1$, the simple cylinder model of Eq.\ref{cyl} has been used; the factor $F_p$, Eq.\ref{F_p}, which determines the fraction of the visible sun, can only obtain values of exactly 0 and 1 in this model. Model two ($s2$) and model three ($s3$) are represented in Eq.\ref{F_p_1} and \ref{F_p_2}, respectively. The factor $F_p$ in the three different shadow models, which have been used,  is illustrated in Fig.\ref{F_p}. Although the difference between the dual cone model $s2$ and the physically adapted model $s3$ is significant for objects in low Earth orbits \cite{hujsak_shadow}, it is small for the objects in the geostationary ring. \\
\\
 Passing through Earth's shadow has not only an effect on the orbit, since the radiation pressure is absent or diminished over parts of the orbit, but also  induces a radiation pressure gradient over the surface of the objects, which leads to a torque.  Fig.\ref{torque} shows the values of the solar torque over the object \textit{PET} for two passages through Earth's shadow, and a close up on the first passage. It can be seen, that for the simply cylindrical model $s1$, no torque is picked up, but it is just integrated over in the propagation, which is not likely to be a good representation of the true physical situation. In the cylindrical model there is only one precise epoch, at which a shadow is cast over the object. In the integration step size of around 10 seconds, this moment at which a  small torque could be picked up is not captured. For the dual cone model $s2$ and the enhanced dual cone model $s3$ comparable torques are picked up. Fig.\ref{euler_shadow} shows the Euler angles in comparison to the initial values for the shadow models $s1$ and $s2$. In those model, during the whole time of the  in Earth's shadow, the radiation is varying over the size of the object. This model is closer to the reality that the object is experiencing in orbit at a passage through Earth's shadow. In accordance with the torques, only small changes in the Euler angles comparable to those in absence of any shadow paths are induced by  Earth's shadow are induced by the shadow passage in model $s1$. The difference between model $s2$ and $s3$ is small, the difference between the Euler angles of the two models is shown in Fig.\ref{s2s3}, it is orders of magnitudes smaller than the total changes, see Fig\ref{euler_shadow}. As a comparison, the same Earth's shadow passages and models are simulated for object \textit{KAP}, the figures can be found in the supplement material. Although the object is heavier comparable changes in the Euler angles at the passages through Earth's shadow are visible. \\
\\
Fig.\ref{eccinc_rel} shows the difference in the eccentricity and inclination of object \textit{PET} for shadow model $s1$ minus $s2$. In the supplement materials difference in eccentricity and inclination for the models and $s3$ minus $s2$ is shown. The eccentricity and inclinations start to differ significantly for model $s1$ compared to the other two models starting at the first of the two Earth's shadow passages. The differences between $s3$ and $s2$ are much smaller, a clear step structure is visible, at each passage through Earth's shadow. For a four day propagation period inclination changes of $2.5\cot10^{-5}$ and $1.3\cot10^{-7}$ in eccentricity are induced by at each passage through Earth's shadow.\\
\\
Object \textit{sKAP} has non-uniform reflection properties on one of its sides. The non-uniform reflection properties induce a significant solar torque, which is about eight orders of magnitudes higher than the torque induced at at Earth's shadow passages, see supplement materials. This leads to a rapid change in all three Euler angles over the propagation period of four days: The first and the third Euler angles show the same beat frequencies with varying amplitudes with periods of about six hours. See supplement materials for Euler angles plots. No full turn is reached in the third body axis, but are more or less in a swaying motion. Fig.\ref{sKAPomega} shows the angular velocities and supports, that the object does not simply spin up over time, but the variations in increasing and decreasing angular velocities in all three axis indicate a swaying motion rather than a rapid spin.\\
\\
Object \textit{rKAP} has the same non-uniform reflection properties as object \textit{sKAP}, but one third of the object is curled up on one end. This leads to a smaller AMR value, but also to a gravitational torque acting on the object, because the geometrical center and the center of mass of the object do not coincide any more. The solar torque is higher than for object \textit{sKAP}, because the center of gravity of the object has shifted not only relative to the geometrical center but also to the center of pressure of the object. In the Euler angles, a beat frequency is visible in the first and the third Euler angle, with a period of about 12 hours.  For plots of the torques and Euler angles please refer to the supplement materials. Fig.\ref{rKAPomega} shows the angular velocity, which shows very rapid changes, but appears in the peak values stable over time. The same beat frequency is visible in the angular velocity which was seen in the two Euler angles. As for object \textit{sKAP} angular velocity is experiencing rapid changes, unlike in the prior case. The rotation around all three body axis, and maximum angular velocity in both, forward and backward direction are reached. Most prominently in the rotation around the third axis, the beat frequency is visible again.\\
\\
Fig.\ref{differ} shows the orbital elements eccentricity and inclination of the object \textit{sKAP} and \textit{rKAP}, as well as the objects \textit{ball-sKAP} an \textit{ball-rKAP}, the canon ball reference objects with the same reflection properties. Schildknecht \cite{Habilthomas} and Scheeres\cite{scheeres_2011} suggest that a rapid rotation of a plate like object averages out and leads to the same effective surface as of a canon ball object. This calculation assumes uniform reflection properties, which is not the case here, because the non-uniform reflection properties are the driver for the rapid attitude motion. Fig.\ref{differ} shows that the orbital evolution of the canon ball model \textit{ball-sKAP} and of the flat plate \textit{sKAP}, as well as of \textit{ball-rKAP} and \textit{rKAP}. For object \textit{sKAP} the canon ball model shows a similar but not identical evolution in inclination, which may be assumed to average out over longer propagation intervals, but the evolution of the eccentricity clearly shows a secular effect which will only increase to differ over time. Object \textit{rKAP} shows an irregular evolution in the inclination, which is introduced by the difference between the center or mass an the geometric center of the object.
\begin{figure}[ht]
  \centering
  \subfloat[]{\includegraphics[width=0.55\textwidth]{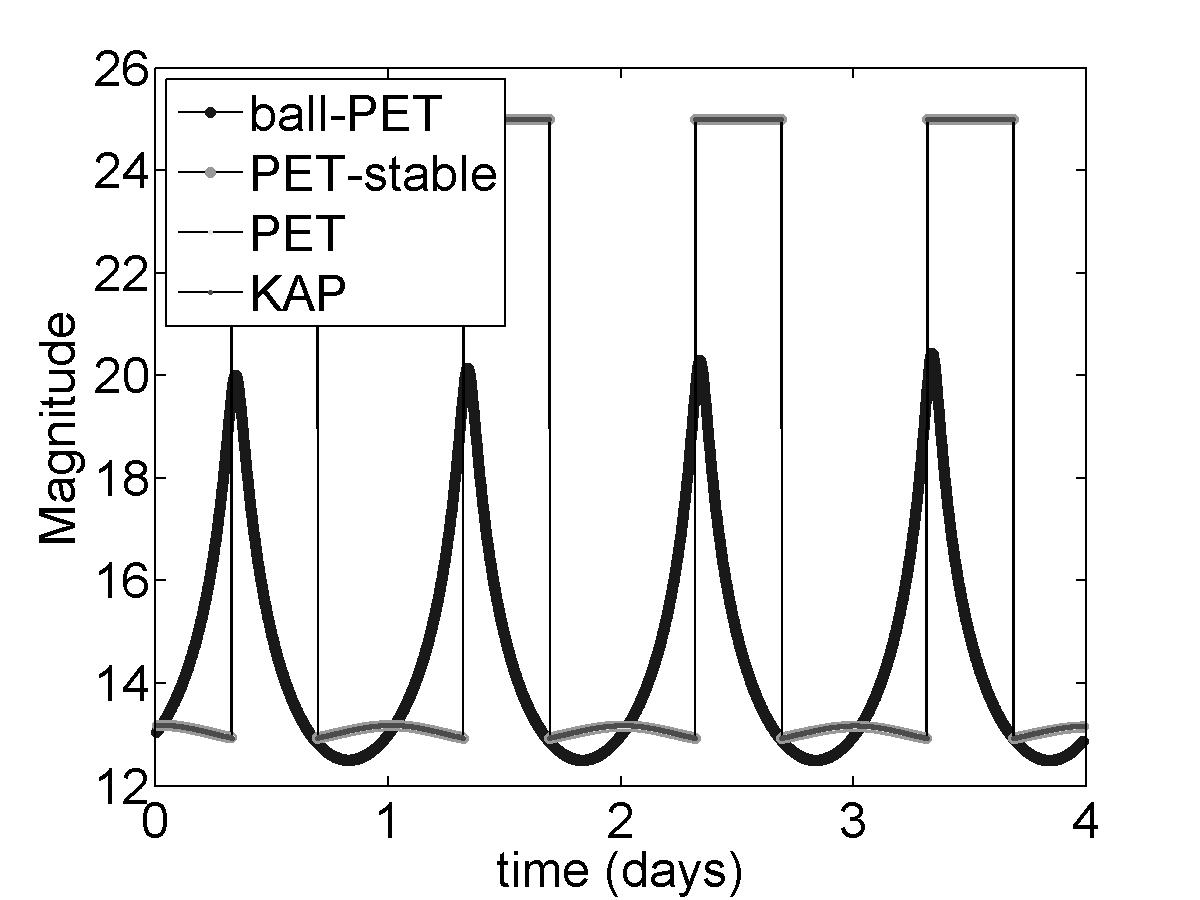}} 
  \subfloat[]{\includegraphics[width=0.55\textwidth, height=5.0cm]{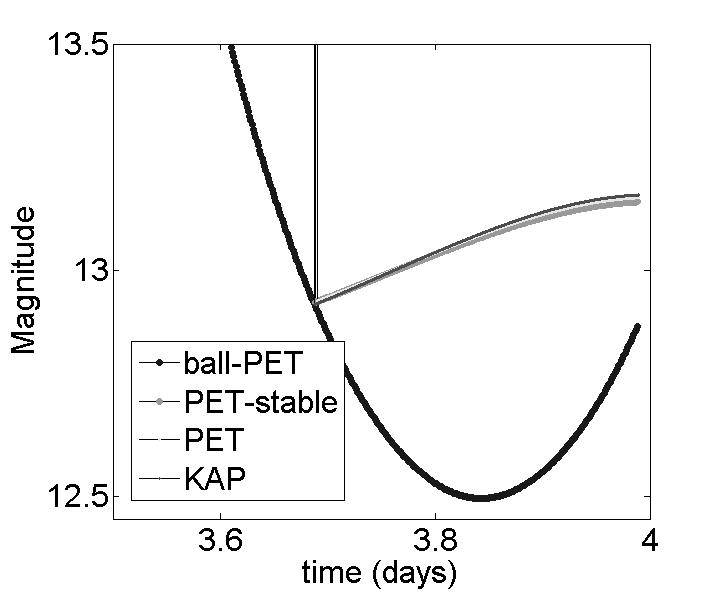}}
  \caption{\bf Simulated light curves without visibility constraints of spherical object \textit{ball-PET}, and rigid plate objects \textit{PET-stable}, with a fixed attitude orientation, as well as object \textit{PET} and object \textit{KAP}.}
  \label{allmag}
\end{figure}
\begin{figure}[h!]
  \centering
  \subfloat[]{\includegraphics[width=0.55\textwidth]{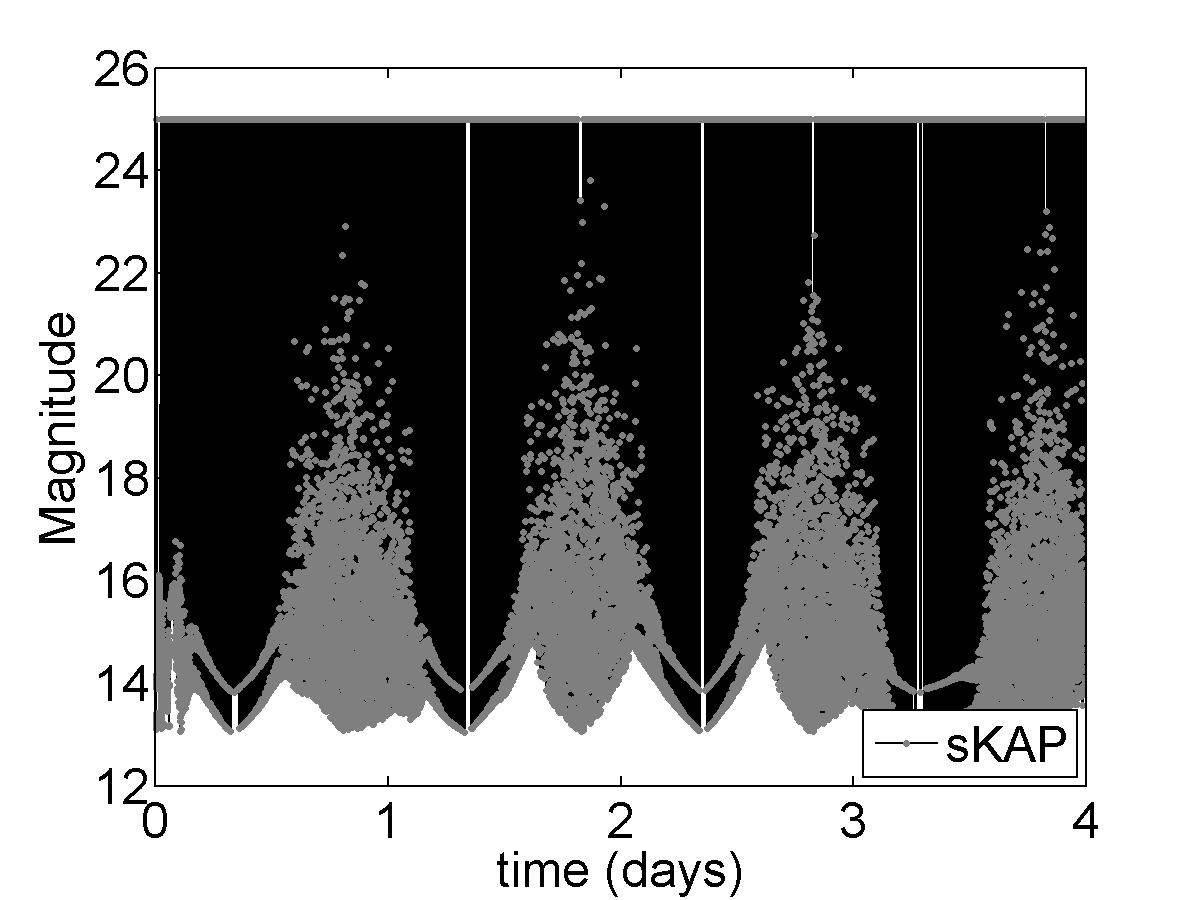}} 
  \subfloat[]{\includegraphics[width=0.55\textwidth]{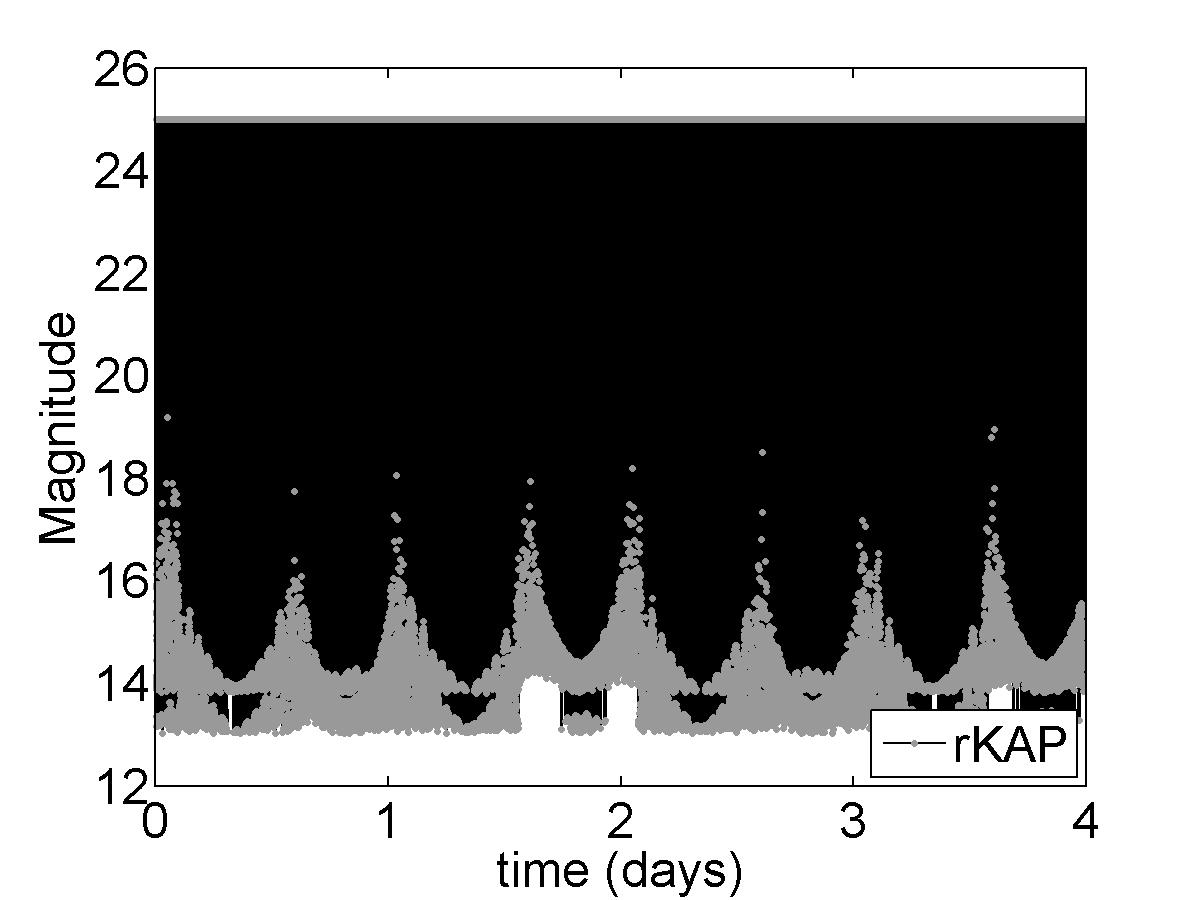}} 
  \caption{\bf Simulated light curves without visibility constraints of rigid plate object \textit{sKAP} and  \textit{rKAP}.}
  \label{lightskap}
\end{figure}
 This leads to even larger differences in the inclination between the object \textit{rKAP} and its spherical counterpart $rKAP-ball$. A similar secular trend as for object \textit{sKAP} can be observed for object \textit{rKAP} in the eccentricity.This leads to the result, that even when a rapid attitude motion is present, apparently, the simple assumption, that it would average out and justify the use of a simple canon ball model again, is not correct for all cases. \\
\begin{figure}[h!]
  \centering
  \includegraphics[width=0.55\textwidth]{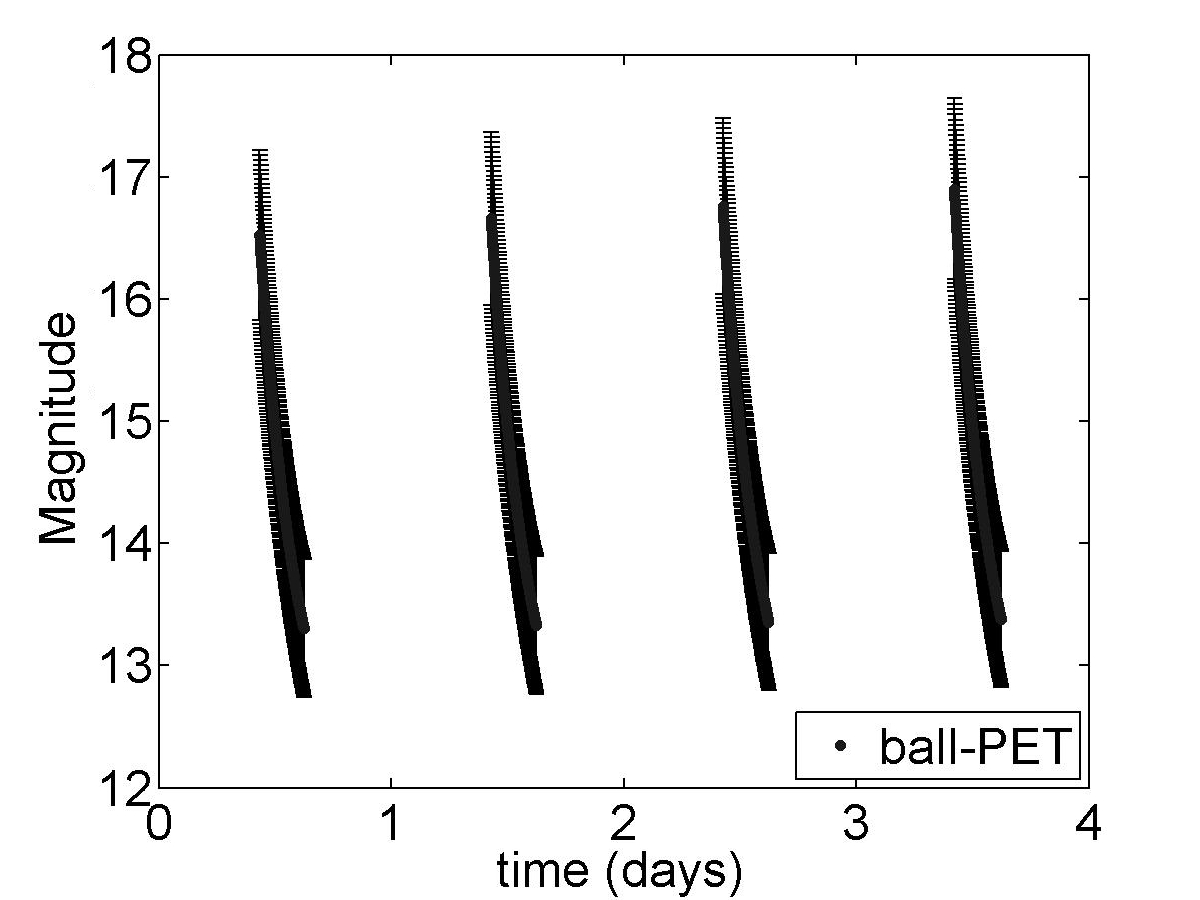}
  \caption{\bf Simulated light curves with visibility constraints and use of a sensor model of spherical object \textit{ball-PET}, and rigid plate objects \textit{PET-stable}, with a fixed attitude orientation, as well as object \textit{PET} and object \textit{KAP}, only object \textit{ball-PET} is visible.}
  \label{SNRallmag}
\end{figure}
\begin{figure}[h!]
  \centering
  \subfloat[]{\includegraphics[width=0.55\textwidth]{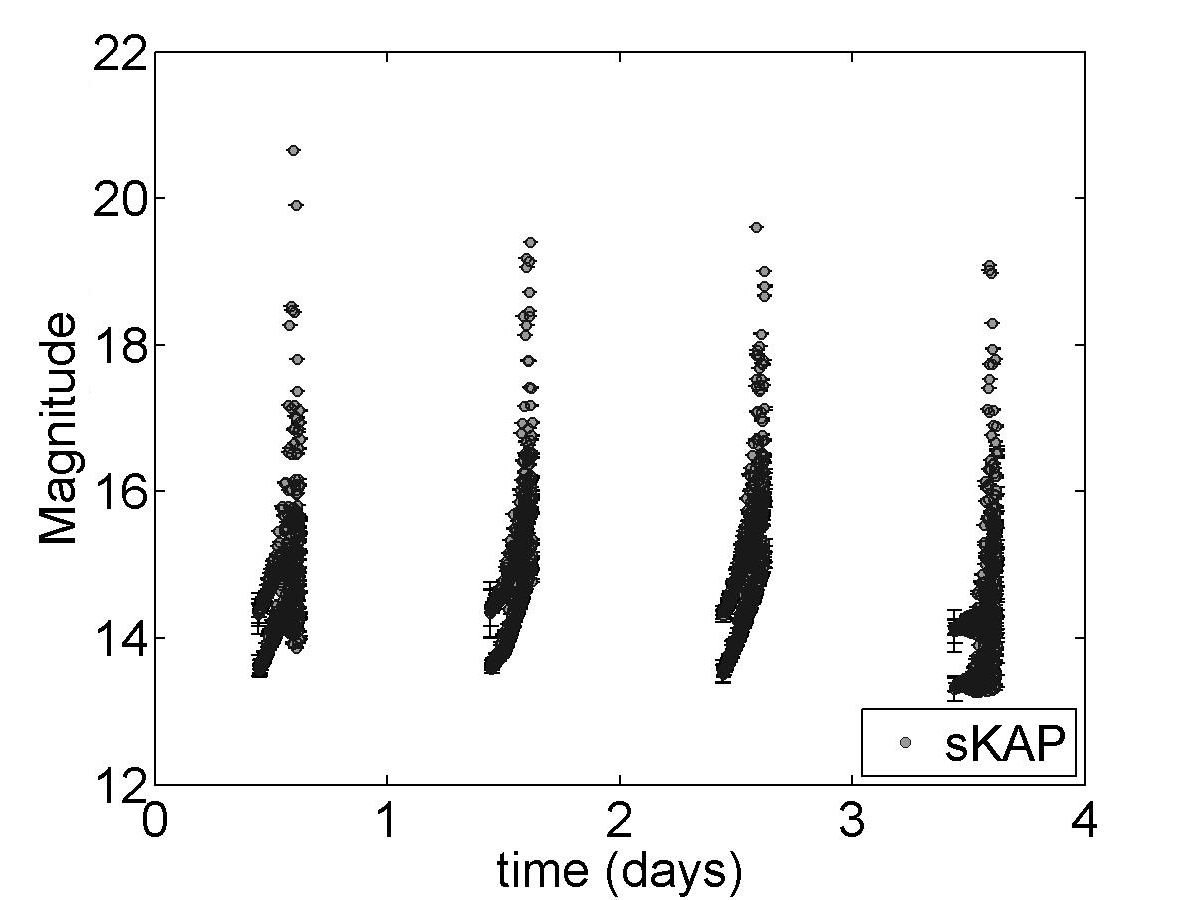}} 
  \subfloat[]{\includegraphics[width=0.55\textwidth]{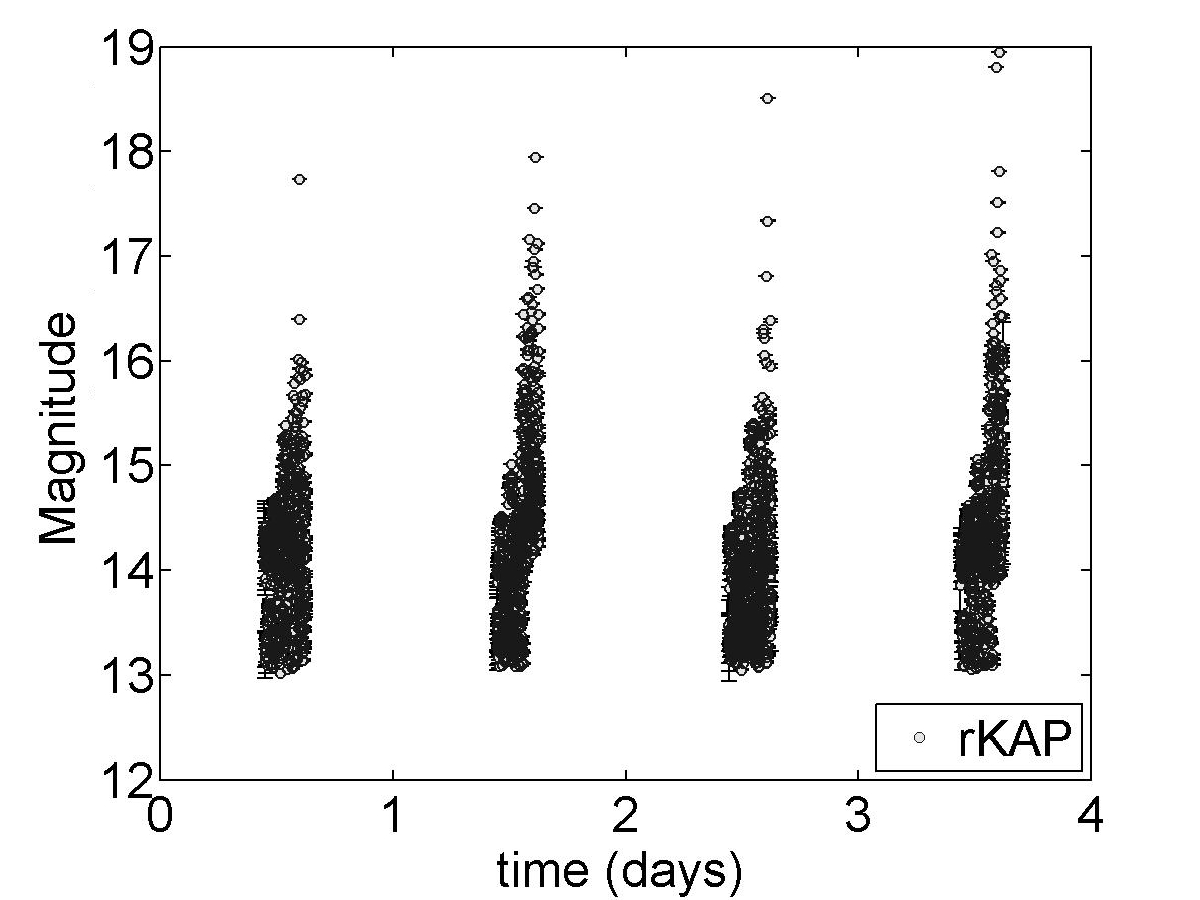}} 
  \caption{\bf Simulated light curves with visibility constraints and use of a sensor model of rigid sheet object \textit{sKAP} and \textit{rKAP}.}
  \label{SNRlightskap}
\end{figure}
\\
The light curves of the simulated objects have been determined as observed from a ground based observer, located in central Europe. In a first step no sensor model was used and not visibility constraints have been applied. As the magnitude of the mean sky background the standard value of 25.0 has been assumed \cite{25mag}, which is equal to no light that is reflected off the object. Fig.\ref{allmag} shows the theoretical light curves of the canon ball object \textit{ball-PET}, the rigid plate object \textit{PET}, once with a fixed attitude \textit{PET-stable} and with a fully six degree of freedom modeled orbit and attitude motion \textit{PET}, as well as the object \textit{KAP}, modeled with coupled orbit and attitude motion. It shows the expected brightness variations of the canon-ball object and the significantly different one of all plate objects. On the large scale, the detected magnitudes of the plate objects do coincide, the differences in magnitudes introduced by thedifferences in attitude before and after the passages through Earth's shadow are of the order of less than one tenth of a magnitude and are therefore practically non-detectable. It is visible that time intervals, in which the plate objects appear very bright, with magnitudes of 13 alternates with phases in which the objects are not visible, magnitude 25. Fig.\ref{lightskap} shows the light curves of the object \textit{sKAP} and object \textit{rKAP}, respectively. In both light curves rapid brightness variation do occur, only very short dark phases are visible in contrary to the case for the stable plates. Every low magnitude is alternated with a measurement, in which the object is not reflecting any light (magnitude 25). The bright data points form for both objects a pattern similar to the reflection of a spherical object, with a period of approximately 12 hours, two such pattern seem to be superimposed in the light curves of both objects. \\\\
In a second step, the sensor model, as established in the previous section has been utilized. The sensor is assumed to have an aperture of one meter in diameter, equipped with a CCD with an quantum efficiency of 0.8, a dark noise of 20 and readout noise of 10 electrons per pixel at standard cooling, a gain of 5.0 with a standard deviation of 0.25, exposure time has been chosen to be three seconds. A pixel scale of two arcseconds per pixel was assumed, leading to the assumption of a standard spread of the object function over four pixels. A mean atmospheric extinction coefficient of 0.25 was assumed. All values, have been averaged with the mean wavelengths of 600nm. Visibility constraint are the local horizon as well as civil twilight. As limiting signal to noise ratio 2.5 has been assumed for detection and evaluation of a data point. \\
\\
Fig.\ref{SNRallmag} shows the light curve with the visibility constraints and magnitude uncertainty according to the determined SNR of the objects \textit{ball-PET}, \textit{PET-stable}, \textit{PET} and \textit{KAP}. Only the spherical object \textit{ball-PET} has visible passes for the chosen sensor, the other objects are not detectable in principle. Only one flank of the total light curve is visible, and the cosine dependency, clearly visible in Fig.\ref{allmag}, appears to be washed out by the given magnitude errors. Fig.\ref{SNRlightskap} shows the light curve as determined with the sensor model of object \textit{sKAP} and object \textit{rKAP}.\\
\\
One can clearly see the limiting magnitude of the sensor, which is around the 19th magnitude, unlike in the simulated cases without further constraints, only parts of the whole light curve is visible and it cannot be determined, if the object is not reflecting light in between the bright measurements, which would indicate a very rapid attitude motion, or is just tumbling slower, because data points beyond the detection limit can simply not be counted. The clear pattern, which were visible in Fig.\ref{lightskap} cannot be readily recognized in the noise light curves of Fig.\ref{SNRlightskap}.

\section{Conclusions}
Objects with high area-to-mass ratios (HAMR) have been simulated. The objects were simulated either as rigid flat sheets of one square meter size or as reference objects with identical area-to-mass ratios and reflection properties, as spherical canon ball objects. The materials of  multi-layer insulation (MLI) materials PET and Kapton has been used in the simulation. MLI materials are very likely candidates for actual HAMR objects. \\
\\
The canon ball models showed smaller amplitudes in the daily change of the inclination but a more rapid secular increase in eccentricity compared to corresponding flat objects, which does not average out. It could be shown that even for sheet shaped objects with uniform reflection properties, attitude is significantly altered due to solar radiation torque over the size of the object, during an propagation interval of four days only, leading to changes in the Euler angles up to the order of $10^{-3}$ degrees. This may lead to significant differences in the long term propagation of HAMR objects, when attitude change is neglected. \\
\\
Three different shadow models have been compared, a simple cylindrical model, a dual con model and an enhanced dual cone model. In contrary to objects in low Earth orbit, the differences between the latter two were found to be small for the high altitude orbits investigated in this paper. The cylindrical model did not pick up the expected increase in solar radiation induced torque when entering or leaving Earth shadow. Using the dual cone models, induced such a torque, which lead to significant alteration of the attitude of the objects at each Earth's shadow passage, in the order of the order of $10^{-2}$ degrees in Euler angles.
\\
Rapid attitude motion of the flat sheets is introduced, when non-uniform reflection properties on one of the surfaces is present. The investigated objects do not simply spin up during the propagation interval but end up in a swaying motion, oscillating back and forth with varying angular velocity rates. This change in attitude motion is even more rapid, if in addition to the non-uniform reflection properties a difference between the center of mass and the geometrical center of the object is introduced, by assuming part of the object has rolled up at one end. The very rapid attitude motion leads to inclination and eccentricity changes different from a stable plate. Comparison with corresponding canon ball objects reveal similar trends in the variation of the inclination in case of non-uniform reflection properties. In case an additional gravitational torque is present, that inclination variations become irregular and do not resemble the variations of the canon-ball object.In both cases the secular trend in eccentricity differs significantly and the difference is increasing over time, which makes a canon ball object not a suitable approximation in the propagation of the investigated flat plate objects. \\
\\
Light curves of the objects have been simulated. The canon ball objects and rigid sheet objects with uniform reflection properties show well distinguishable brightness pattern. The change in attitude motion introduced by passages through Earth's shadow lead only to very small magnitude changes, which are not readily detectable. The objects with non-uniform reflection properties, which are in a very rapid attitude motion, show rapid brightness changes, but also superpositions leading to beat frequencies of around 12 hours. A sensor model determining magnitude uncertainties and visibility constraints has been utilized. Using the sensor model, several flat plate objects with  no or small attitude changes, have not been observable any more, due to the long paths, in which no light is reflected towards the observer and local visibility constraints. The spherical object has still been observable, as well as the flat plates in rapid attitude motion. The beat frequency was not readily to be determined any more, due to the estimated magnitude errors, and because of the effect of the limiting magnitude of the simulated sensor.

\section{Acknowledgment}
The main author would like to thank Thomas Schildknecht for useful discussions. The work of the main author was supported by the National Research Council. 

\bibliographystyle{apalike}  
\bibliography{frueh}

\end{document}


\title{Coupled Orbit-Attitude Dynamics of High Area-to-Mass Ratio (HAMR) Objects: Influence of Solar Radiation Pressure, Earth's Shadow and the Visibility in Light Curves
}


\author{Carolin Fr\"uh          \and
        Thomas M. Kelecy        \and Moriba K. Jah}


\institute{C. Fr\"uh  \at
              NRC Postdoc, Air Force Research Laboratory, Space Vehicles Directorate, Research Assistant Professor, University of New exico, Albuquerque, USA \\
              Tel.: +1-505-277-2761 \\
              Fax: +1-505-277-1571\\
              \email{carolin.frueh@gmail.com}           
           \and
           Thomas M. Kelecy \at
              The Boeing Company, 5555 Tech Center Dr., Ste 400, Colorado Springs, CO 80919
	\and 
	Moriba K. Jah \at Air Force Research Laboratory, Space Vehicles Directorate, NM, 87117, USA
}

\date{Received: date / Accepted: date}

\maketitle

\section{Supplement Materials}
\subsection{Object Materials and Properties}
\begin{table}[h!]
 \begin{center}
  \begin{threeparttable}
   \caption{Fabrication values of MLI pieces \cite{redbook}: Thickness [$mm$], Coating, AMR value  [$m^2/kg$], reflection (specular and diffuse) and absorption values ($C_{\mathrm{s}}$, $C_{\mathrm{d}}$, $C_{\mathrm{a}}$) \cite{redbook}.}
   \label{MLI}
   \begin{tabular}{lll}
    \bf{PET  [$0.25mm$]}&
    AMR [$m^2/kg$] &
    $C_{\mathrm{s}}$, $C_{\mathrm{d}}$, $C_{\mathrm{a}}$\\
\hline
    coated & 111.11 &  0.60 0.26 0.14 \\  
\hline
\bf{Kapton  [$1mm$]}&&\\
    coated  & 26.30 & 0.60   0.26   0.14 \\ 
    uncoated & 26.30 &  0.00   0.10   0.90 \\
   \end{tabular}
  \end{threeparttable}
 \end{center}
\end{table}
Double aluminum coated PET$^{\textregistered}$ and single coated Kapton$^{\textregistered}$ has been simulated, the properties of the these MLI materials are listed in Tab.\ref{MLI}.  $C_{\mathrm{s,d,a},i}$ are the coefficients for specular, diffuse reflection.\\
\begin{table}
 \begin{center}
  \begin{threeparttable}
   \caption{Reference canon ball objects.}
   \label{balls}
   \begin{tabular}{lll}
\bf Name & \bf AMR [$m^2/kg$]  & \bf{$C_{\mathrm{s}}$, $C_{\mathrm{d}}$, $C_{\mathrm{a}}$}\\
\hline
    ball-PET & 111.11 &  0.60 0.26  0.14  \\ 
    ball-PETeff & 25.40 &  0.60 0.26  0.14  \\ 
    ball-sKAP  & 26.30 & 0.20   0.15   0.65 \\ 
    ball-rKAP & 19.70 &  0.20   0.15   0.65 \\
   \end{tabular}
  \end{threeparttable}
 \end{center}
\end{table}
\subsection{Supplements to the Simulation Results}
\begin{figure}[h!]
  \centering
  \subfloat[]{\includegraphics[width=0.5\textwidth]{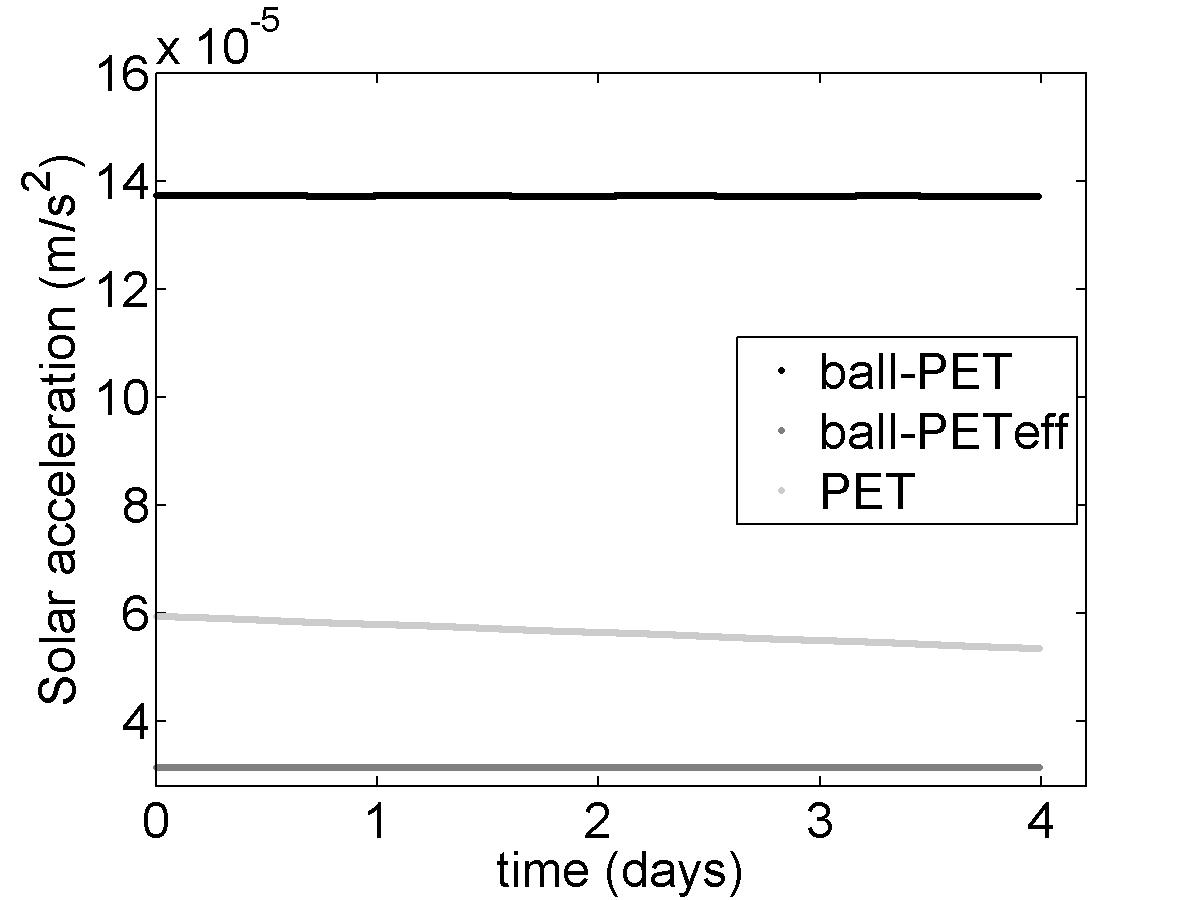}} 
  \subfloat[]{\includegraphics[width=0.5\textwidth]{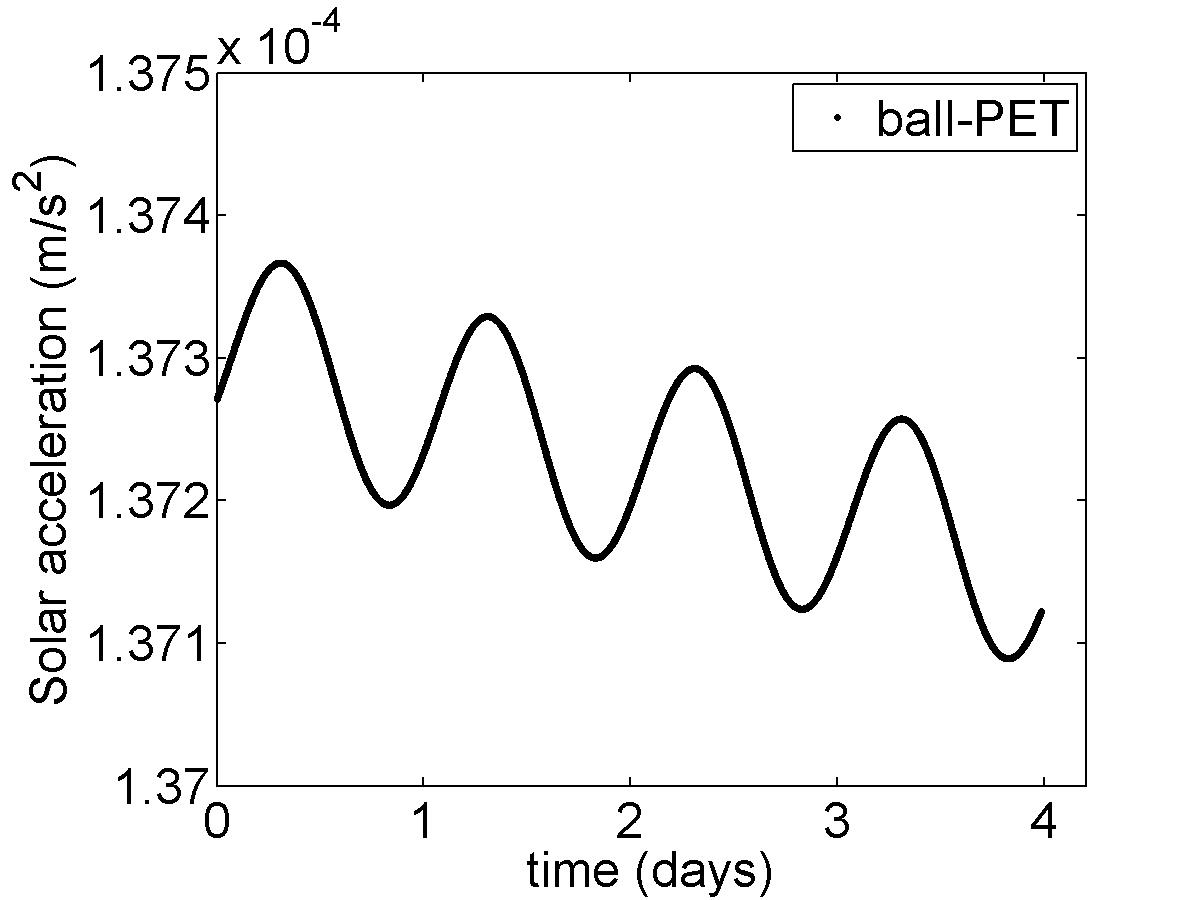}}
  \caption{\bf (a) Acceleration due to direct solar radiation pressure  \textit{ball-PET}, \textit{ball-PETeff}, \textit{PET}. (b) close up on the acceleration of the \textit{ball-PET} object.}
  \label{pet0acc}
\end{figure}
\begin{figure}[h!]
  \centering
  \subfloat[]{\includegraphics[width=0.5\textwidth]{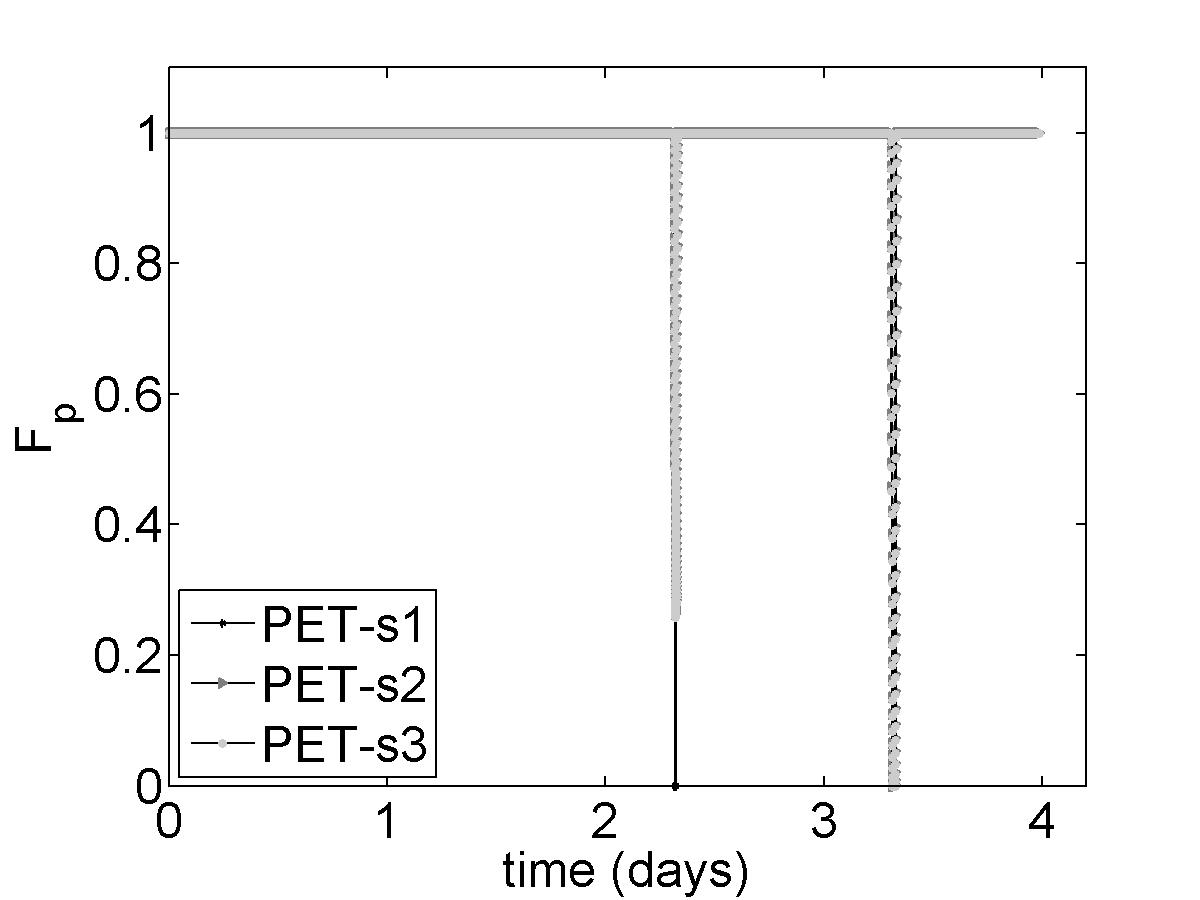}} 
  \subfloat[]{\includegraphics[width=0.5\textwidth]{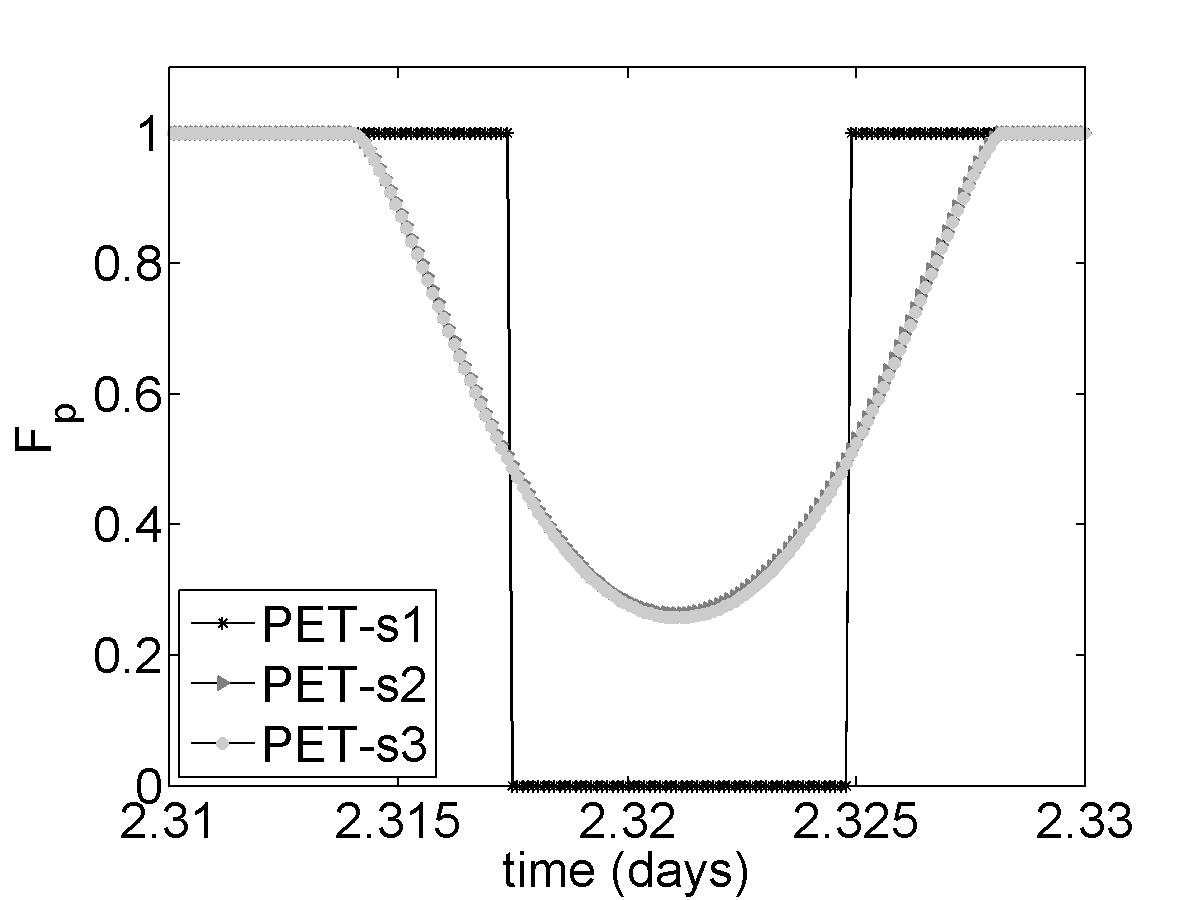}}
  \caption{\bf Fraction of the visible sun $F_p$, for different shadow models: $s1$, $s2$, $s3$ l (a) two shadow passes at 2.3 and 3.3 days, (b) zoom on the first passage through shadow.}
  \label{F_p}
\end{figure}
\newpage
\begin{figure}[h!]
  \centering
  \subfloat[]{\includegraphics[width=0.45\textwidth]{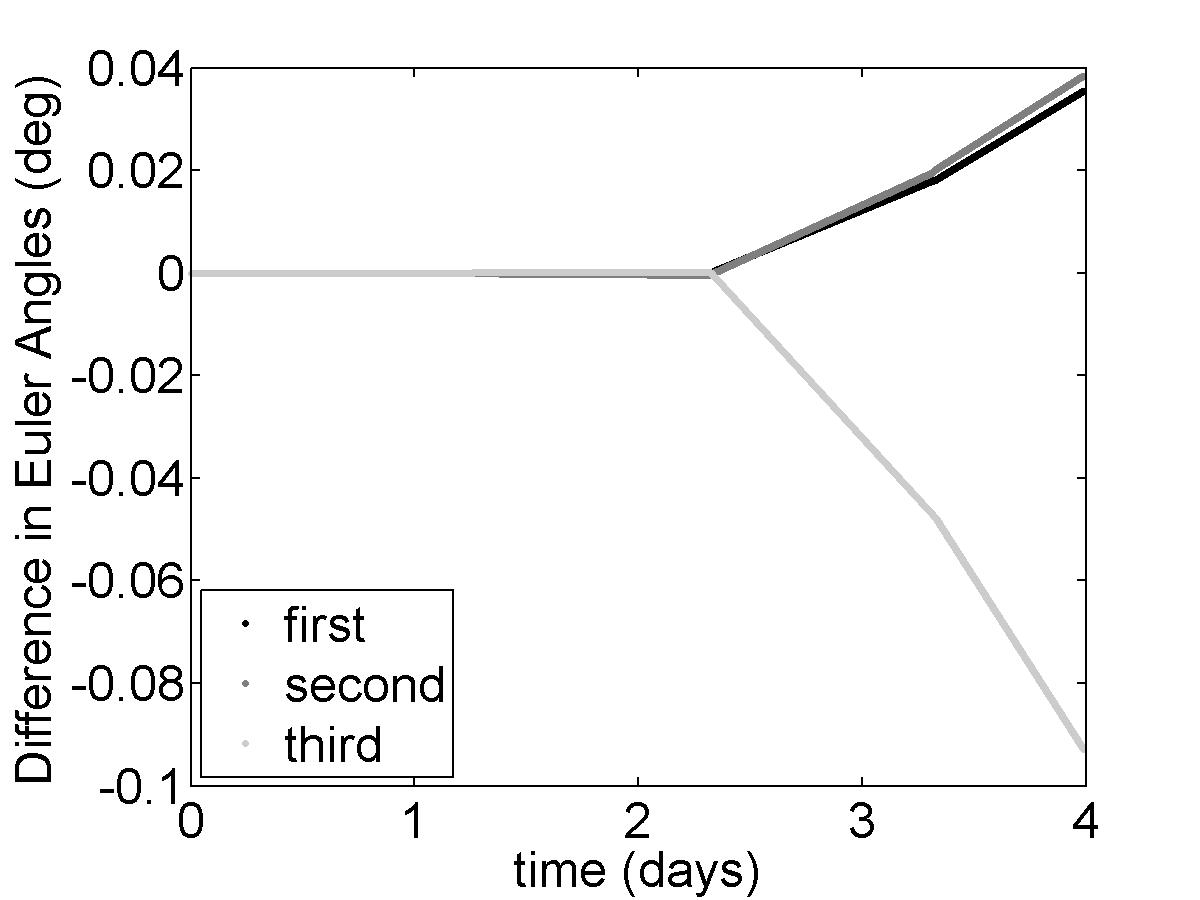}} 
  \subfloat[]{\includegraphics[width=0.45\textwidth]{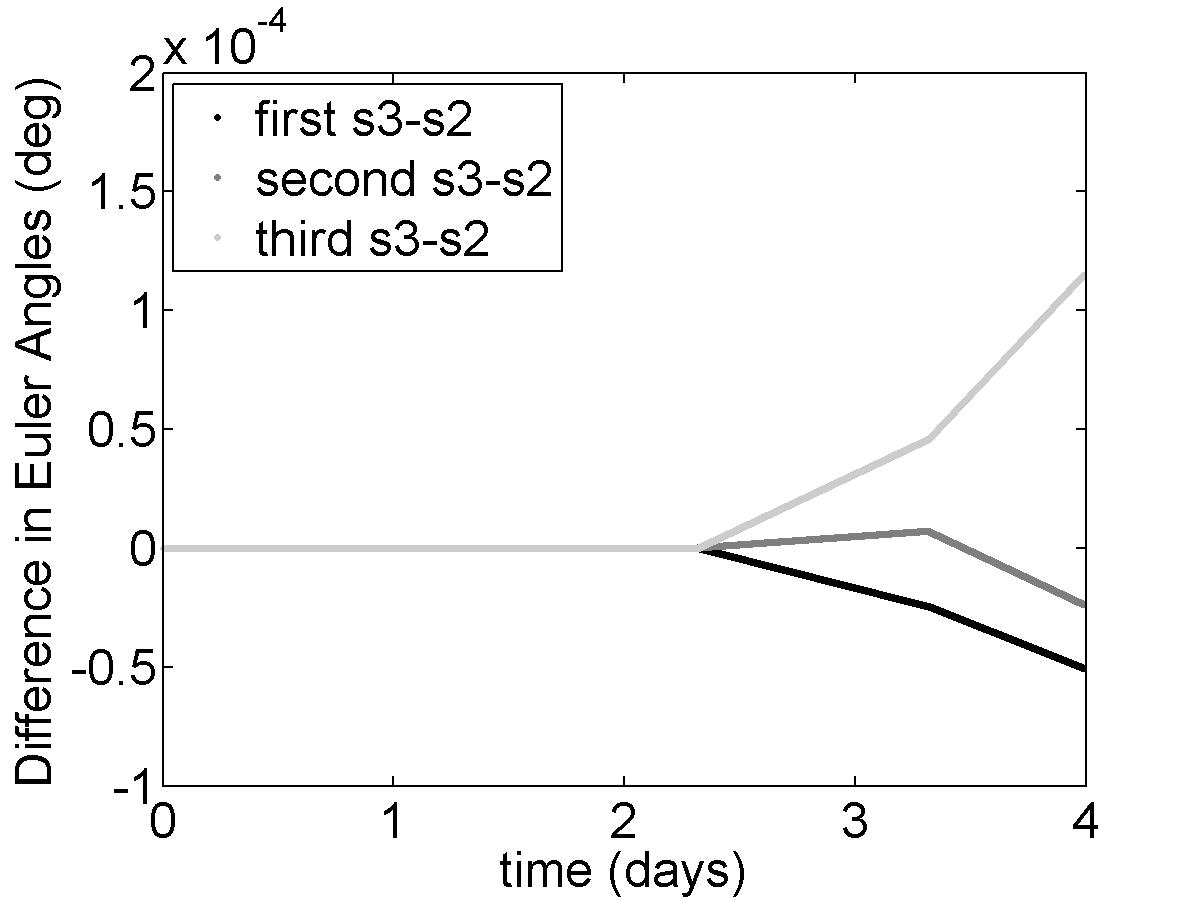}}
  \caption{\bf First, second and third Euler angle evolution of object \textit{KAP} relative to the initial values for (a) shadow model $s2$ and (b) for the difference of models $s3$-$s2$.}
  \label{euler_shadow_kap}
\end{figure}
\begin{figure}[h!]
  \centering
 \subfloat[]{\includegraphics[width=0.5\textwidth]{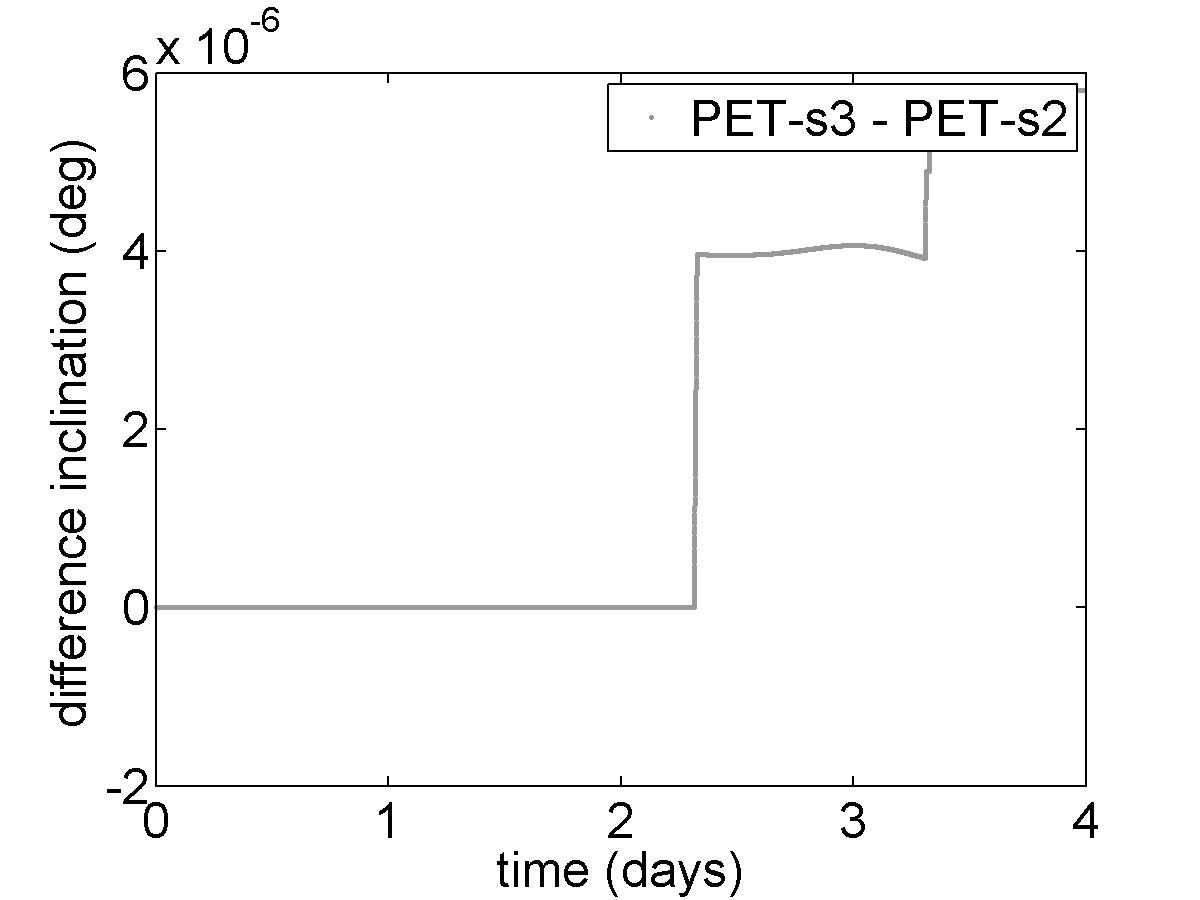}}
  \subfloat[]{\includegraphics[width=0.5\textwidth]{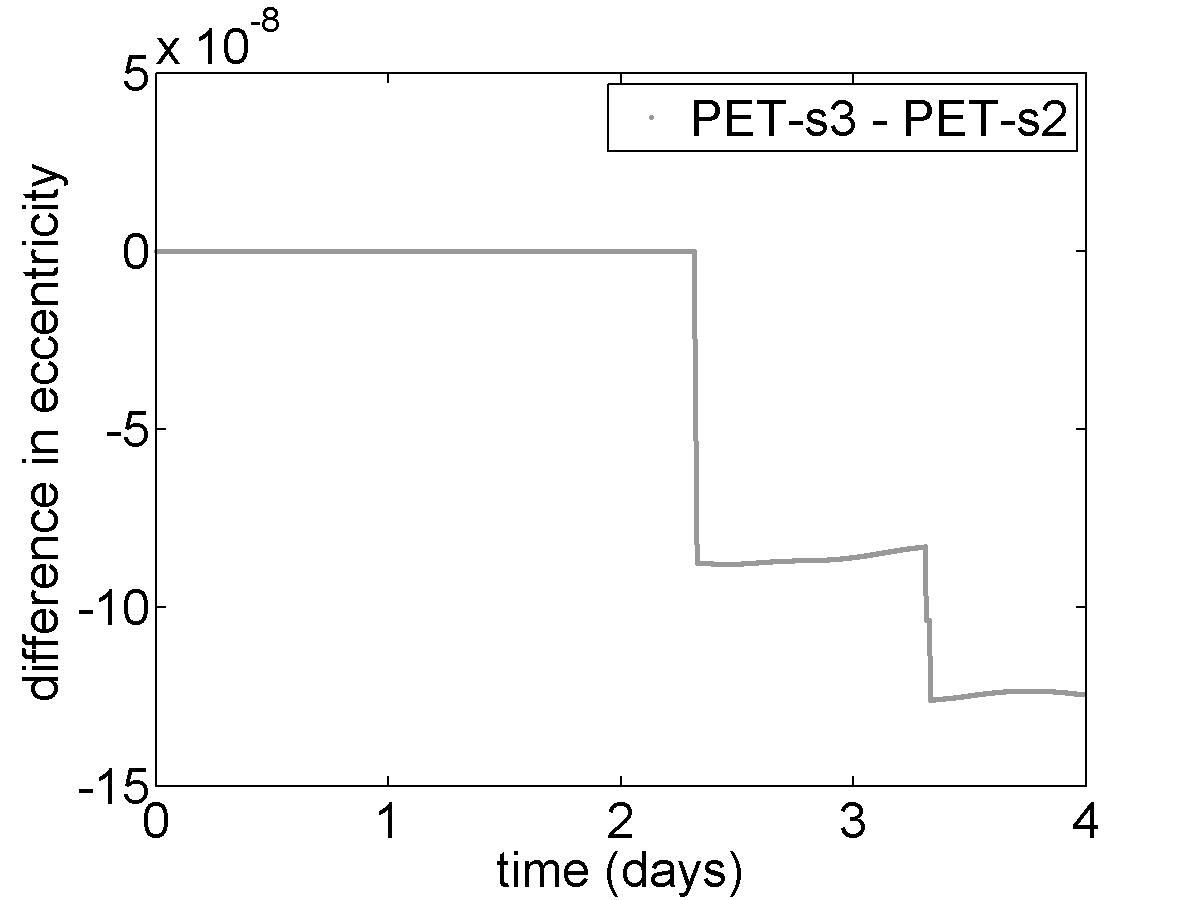}}
  \caption{\bf Differences in the inclination and eccentricity evolution of object \textit{PET} between model $s2$ and model $s3$.}
  \label{ecc_rel}
\end{figure}
\newpage
\begin{figure}[h!]
  \centering
  \includegraphics[width=0.5\textwidth]{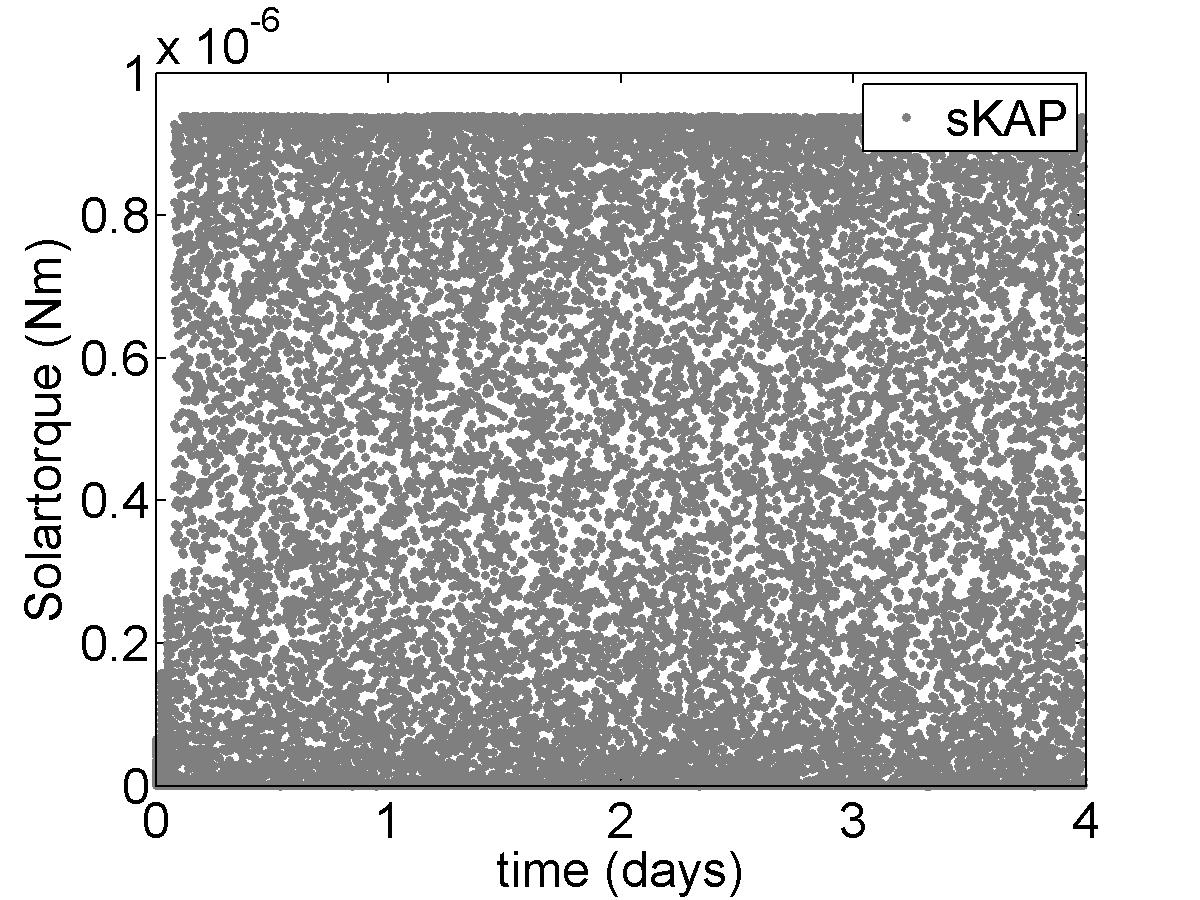}
  \caption{\bf Solar torque induced by non-uniform reflection properties of object \textit{sKAP}.}
  \label{sKAPtorq}
\end{figure}
\begin{figure}[h!]
  \centering
  \subfloat[]{\includegraphics[width=0.5\textwidth]{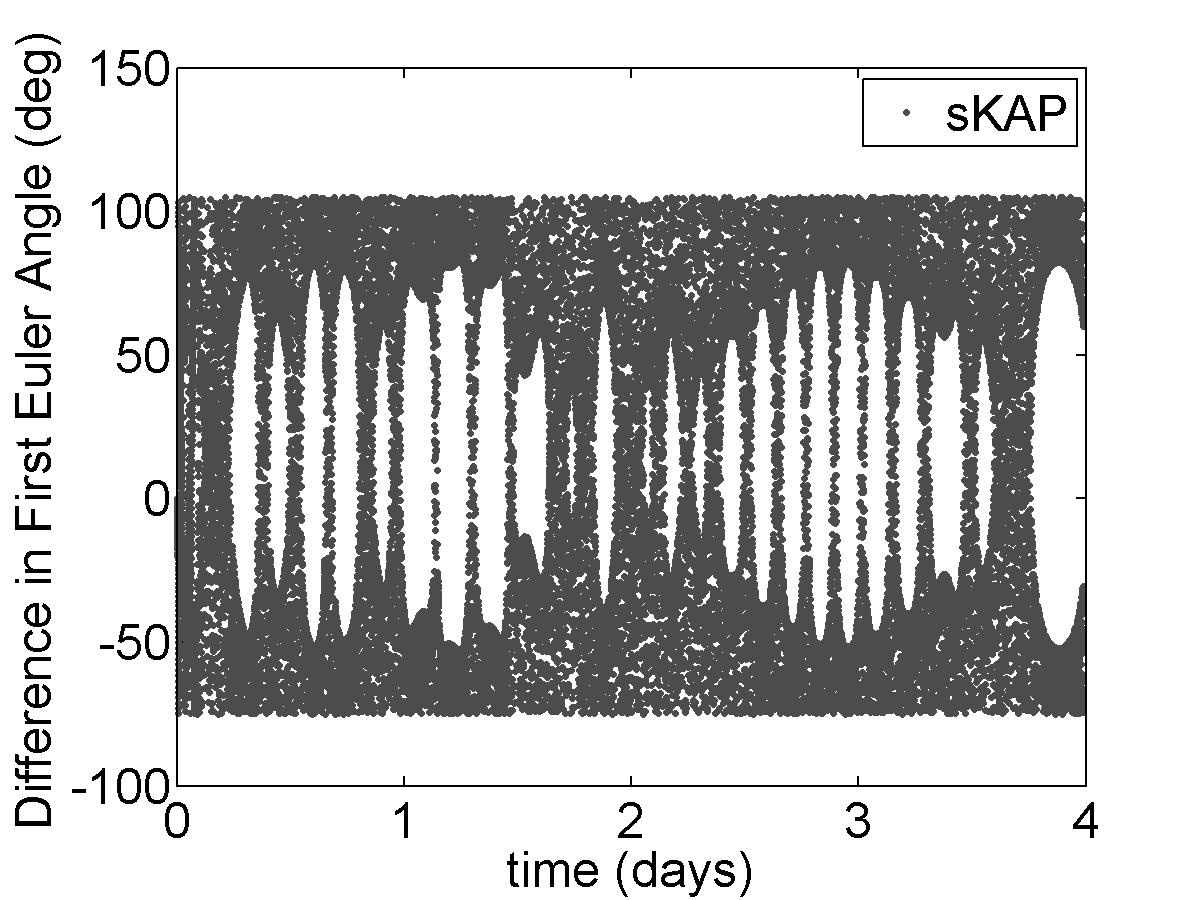}} 
  \subfloat[]{\includegraphics[width=0.5\textwidth]{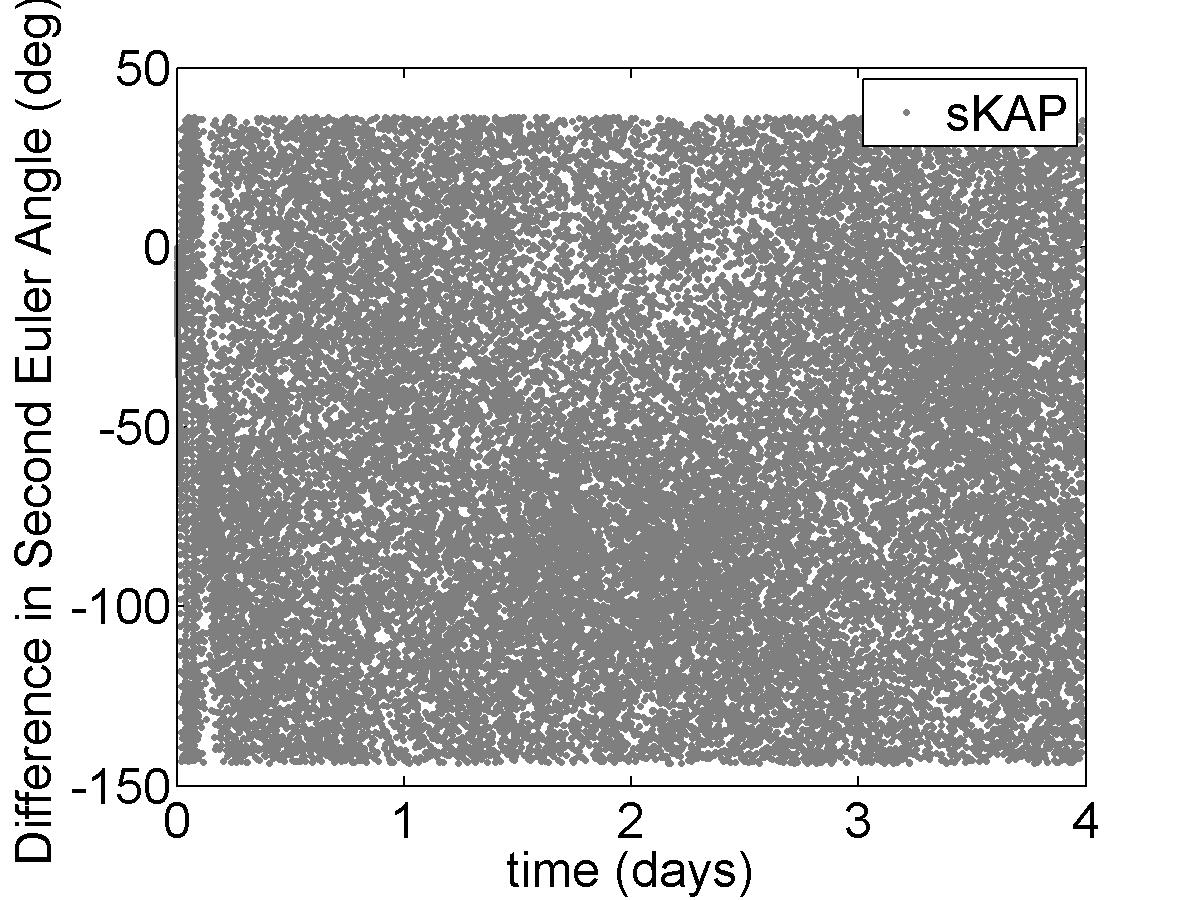}}\\
  \subfloat[]{\includegraphics[ width=0.5\textwidth]{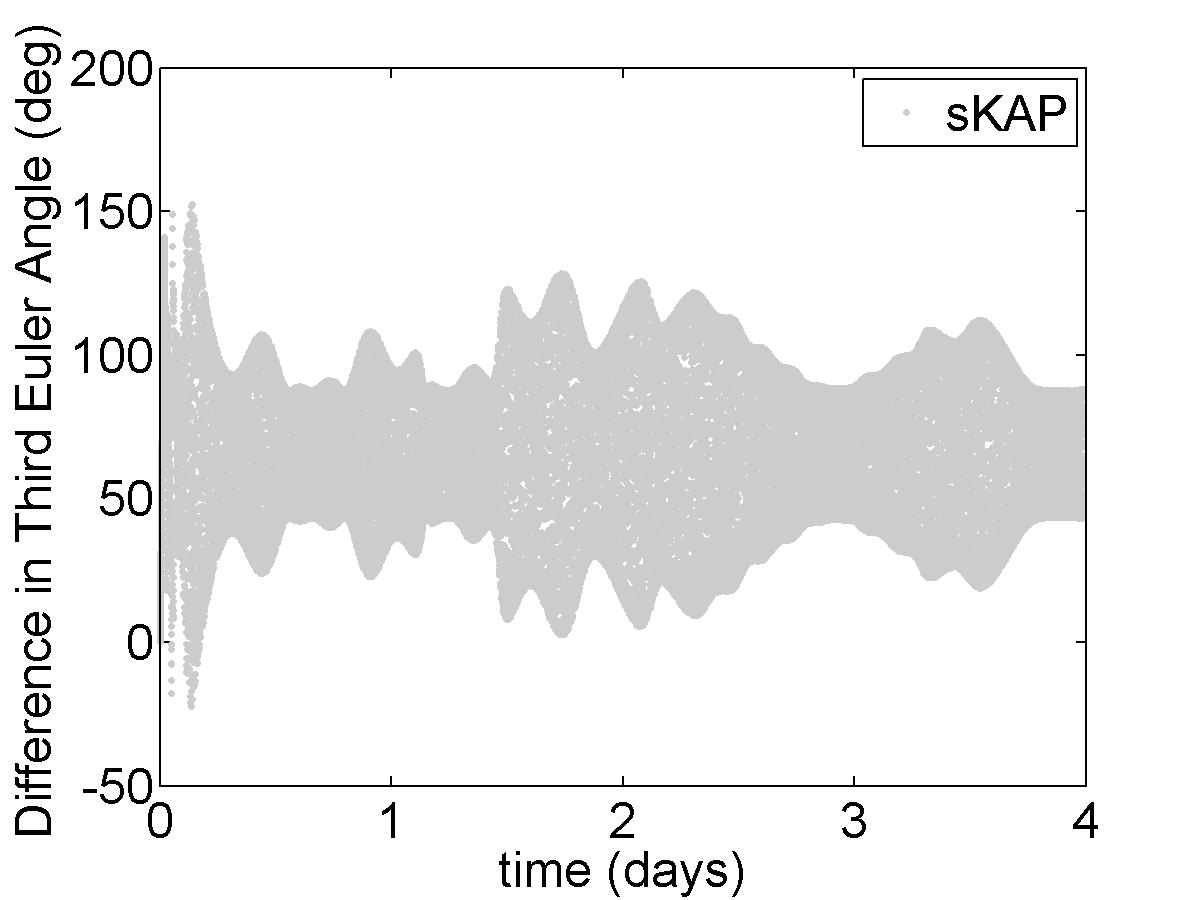}}
  \caption{\bf Euler angle evolution of object \textit{sKAP} relative to the initial values.}
  \label{sKAPeuler}
\end{figure}
\begin{figure}[h!]
  \centering
  \subfloat[]{\includegraphics[ width=0.5\textwidth]{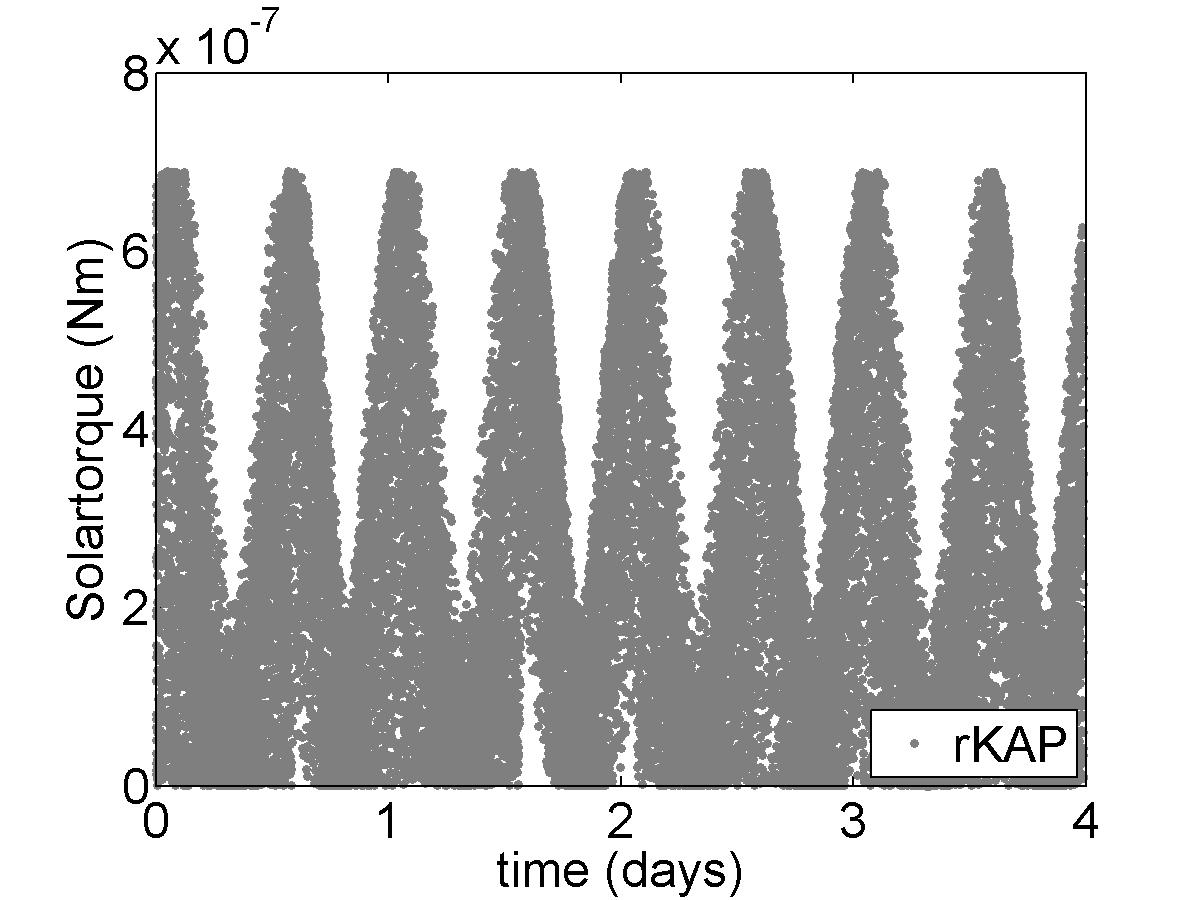}} 
  \subfloat[]{\includegraphics[ width=0.5\textwidth]{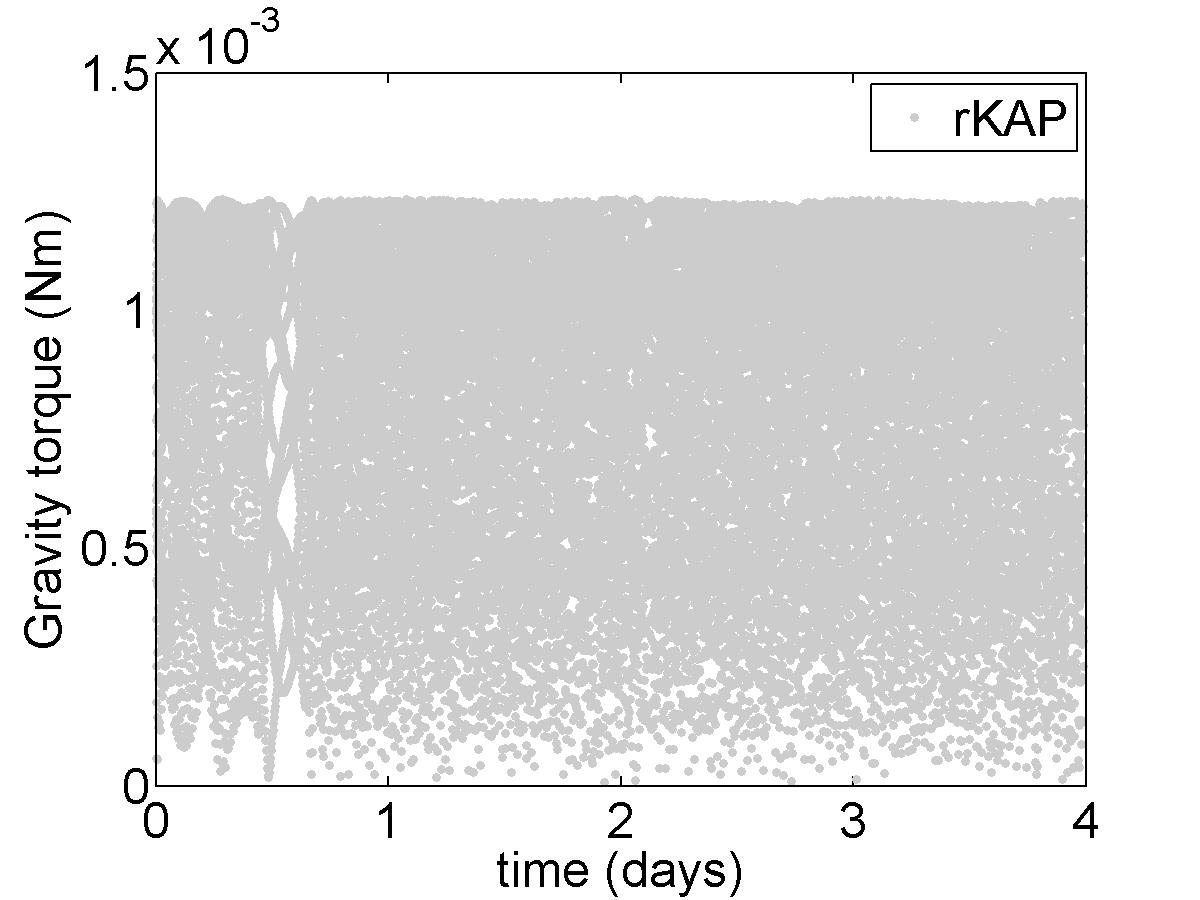}}
  \caption{\bf Solar and gravity torque induced by non-uniform reflection properties as well as center of gravity offset of object \textit{rKAP}.}
  \label{rKAPtorques}
\end{figure}
\begin{figure}[h!]
  \centering
  \subfloat[]{\includegraphics[ width=0.5\textwidth]{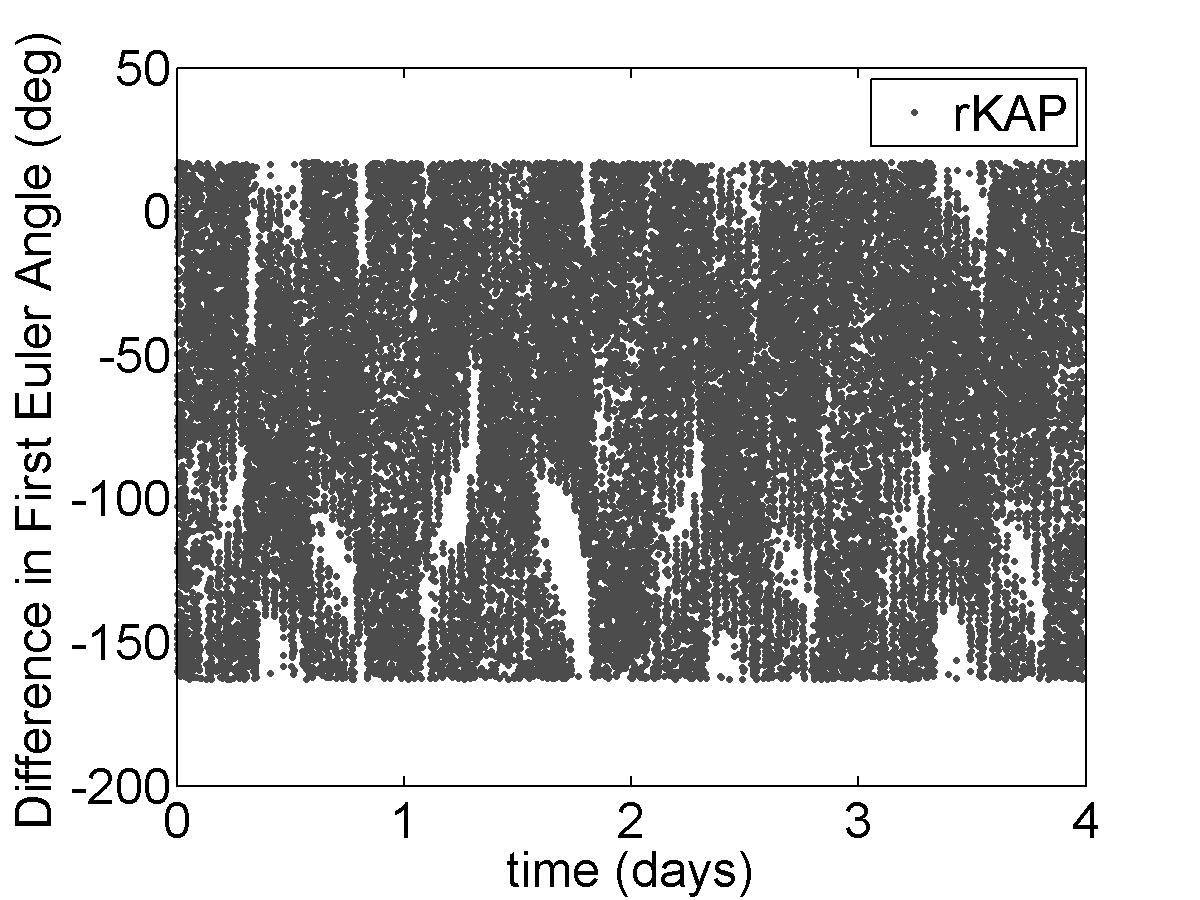}} 
  \subfloat[]{\includegraphics[ width=0.5\textwidth]{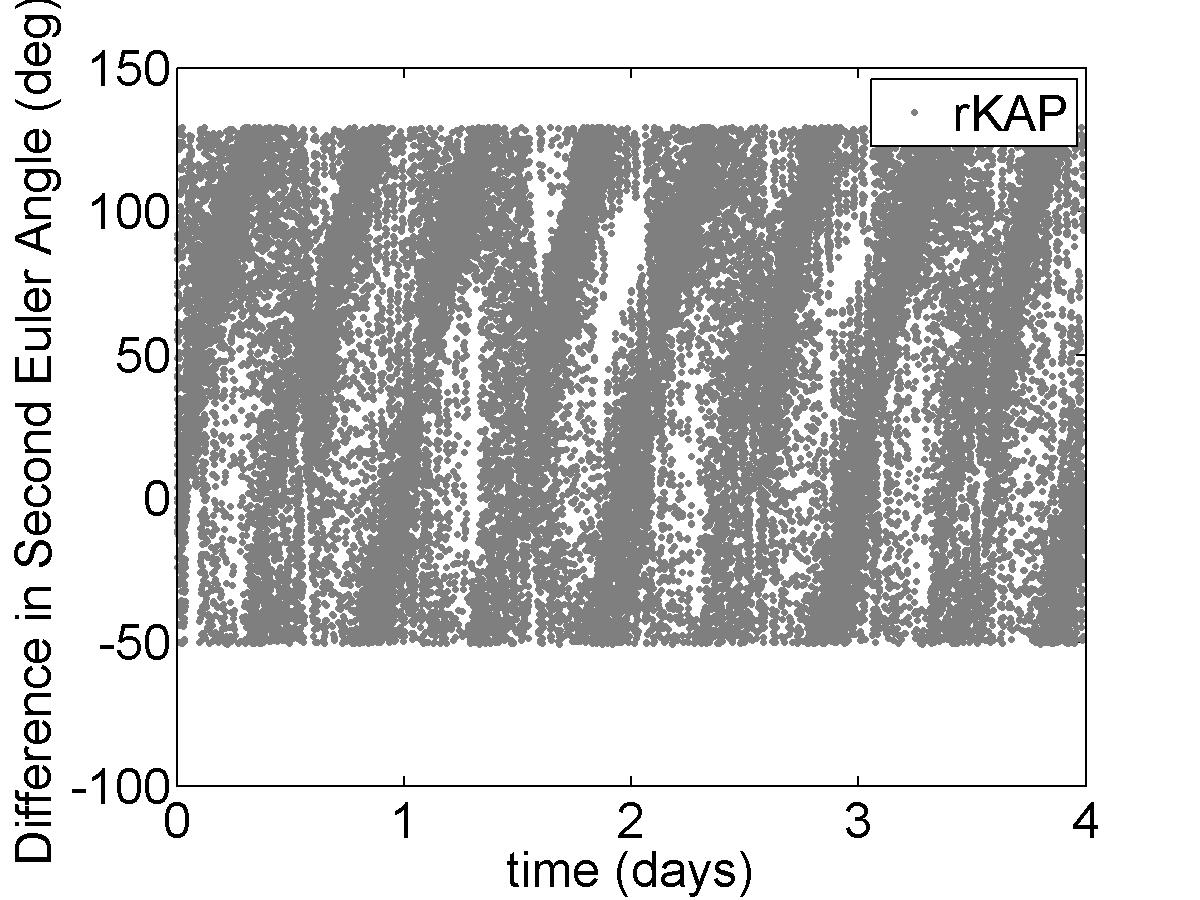}}\\
  \subfloat[]{\includegraphics[ width=0.5\textwidth]{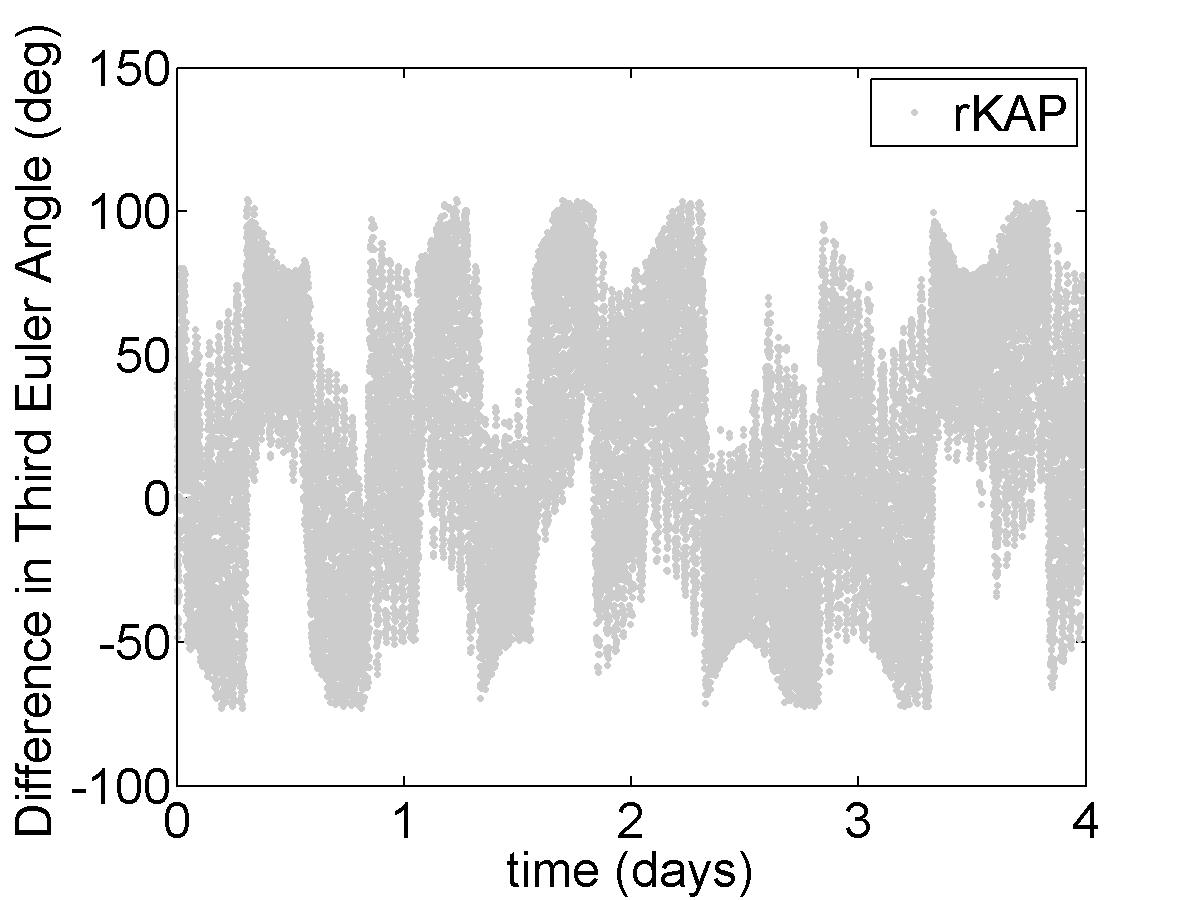}}
  \caption{\bf Euler angle evolution of object \textit{rKAP} relative to the initial values.}
  \label{rKAPeuler}
\end{figure}
\newpage
\bibliographystyle{apalike}  
\bibliography{frueh}